\documentclass[11pt]{article}
\usepackage[utf8]{inputenc}
\usepackage{array}
\usepackage[T1]{fontenc}
\usepackage{tabu}
\usepackage{mathtools}
\usepackage{bm}
\usepackage{authblk}
\usepackage{amsfonts}
\usepackage{color}
\usepackage{tabu}
\usepackage{colortbl}
\usepackage{hyperref}

\title{Opinion polarisation in social networks driven by cognitive dissonance avoidance}

\author[1]{Zoltán Kovács}
\author[1, 2, 4, +]{Anna Zafeiris}
\author[3,1,+,*]{Gergely Palla}

\affil[1]{Department of Biological Physics,  E\"otv\"os University, P\'azm\'any P. stny 1/A, 1117, Budapest, Hungary}
\affil[2]{MTA-ELTE Lendület "Momentum" Innovation Research Group,  E\"otv\"os University, Institute of Archaeological Sciences, Budapest, 1088, Hungary}
\affil[3]{Health Services Management Training Centre, Semmelweis University, K\'utv{\"o}lgyi \'ut 1., 1125 Budapest, Hungary}
\affil[4]{MTA-ELTE "Lendület" Evolutionary Genomics Research Group, P\'azm\'any P. stny 1/A, 1117, Budapest, Hungary}

\affil[*]{gergely.palla@emk.semmelweis.hu}
\affil[+]{these authors contributed equally to this work}

\begin{document}

\maketitle

\begin{abstract}
As the consequences of opinion polarization effect our everyday life in more and more aspect, the understanding of its origins and driving forces becomes increasingly important.
Here we develop an agent-based network model with realistic human traits: individuals in our simulations are endowed with an internal belief system which they attempt to keep as coherent as possible. This desire -- to reassure existing attitudes while avoiding cognitive dissonance -- is one of the most influential and widely accepted theories in social psychology by now. Our model shows that even in networks that start out completely uniform (from a society of clones), this effort leads to fragmentation and polarization, reflected both by the individual beliefs (attitudes) and the emerging communities in the social network. By fine-tuning two parameters: (i) "dissonance penalty", measuring the strength with which agents attempt to avoid cognitive dissonance, and (ii) "triadic closure affinity", the parameter reflecting agents' likelihood to connect with friends of friends, a wide range of possible community structures are observed. 
\end{abstract}

\section*{Introduction}

Network science provides a general approach for the study of complex systems on an extreme wide range, from chemical reactions within living cells through genetic regulation or neural interactions up to the scale of the Internet, the world trade or the global pandemic spreading of diseases\cite{Laci_revmod,Dorog_book,Newman_Barabasi_Watts,Jari_Holme_Phys_Rep,Vespignani_book}. One of the fields within network science of high interdisciplinary interest is the study and the modelling of the structure and dynamics of social networks, where nodes represent individuals and links indicate acquaintances or friendship relations. In general, the formation of social ties in such systems can be affected by many different factors such as the position and movement of people in the physical world (e.g., moving to a new town likely comes with the establishment of new connections) or the movement of people between institutions, firms, etc. (e.g., entering the university also helps finding new friends). However, a further key factor driving the formation of new ties and the deletion of existing links is the opinion or the beliefs of the individuals, with people usually preferring to be connected with others having similar beliefs\cite{Connected09, HomophSoc}.

By the mid-20\textsuperscript{th} century, a new scientific discipline emerged  focusing on exploring and simulating the process through which beliefs and opinions spread within human societies\cite{DeGroot74, Axlrd, TCSch78, SG08}. This area of study, rooted in statistical physics, mathematics and computer science\cite{AlapOpDynOf,peralta2022opinion,PerraSciRep,Sirbu2017,ContOpDynSurvey,SGRev}, became known as \textit{opinion dynamics}, and by now integrates aspects of various fields, such as psychology\cite{CsanyiHied, Scout, KnowldgIllsn, BelBrain, CogPsyTK}, biology\cite{SelfishGene, TC1998} or political sciences\cite{Converse1964, Rokeach63}.  

Since the birth of this discipline in the mid-20\textsuperscript{th} century, the suggested models become more and more refined and realistic regarding both the representation of agents and the social network by which they interact\cite{AlapOpDynOf, g11040065}. The first models highlighted some similarities between the Ising model and the way people alight their opinions\cite{galam1982sociophysics} (agents' opinions being the spins which are pushed to be aligned with their neighbours'). Some other early models suggested similar framework, but with continuous opinions (taking values from a certain, pre-defined interval, usually between $0$ and $1$ or $-1$ and $1$) instead of binary ($+1$/$-1$, yes/no, etc) states \cite{stone1961opinion, chatterjee1977towards, 8ae4e1e508ad490382855c2970e950b7, ContOpDynSurvey}. In general, these models predict global consensus\cite{Olle14} and polarization appears only due to some extra stipulations, such as threshold of communication\cite{OrigDeffuant} or distancing\cite{AxelrodUjPolar}. 
Later, in a seminal paper, direct psychological properties have been also suggested to be incorporated into the models\cite{Latan1981ThePO}, making a big step forward more realistic models. Other approaches proposed to extent the one-scalar representation of agents into more dimensional vectors\cite{AxelrodCikk} which has motivated new directions, among others, studies related to the emerging and co-evolving social networks\cite{RADUCHA2017427}.


In the present work we follow the common approach of combining networks with agent-based simulations: nodes represent agents and the links reflect the social ties between them. However, as a major depart from other models, we propose a biologically and psychologically more realistic representation for the inner state of the agent, being equivalent to a (relatively small) weighted graph between nodes corresponding to concepts, as illustrated in Fig.\ref{fig:trianglesCombined}a. The interactions in the outer social network with other agents can modify the connection weights in this inner representation of the belief system of the agent. This is inspired by a process called "associative learning", referring to the observation that the fundamental learning process in humans -- and in animal as well -- is based on creating new associations between already existing concepts (which process, of course, can be supplemented by incorporating new concepts as well)\cite{AssocLBk, AssocL19, 8945152}. A further important driving force in our model is \textit{cognitive dissonance avoidance}, a widely accepted theory in cognitive psychology\cite{CDL:Cooper07, CDL:MedicalNewsT}, by which agents modify their attitudes towards various concepts and beliefs --  represented by signed node-weights in the inner state representation -- in a way that their beliefs and attitudes remain as coherent as possible (contradiction-free). Lastly, similarly to the majority of evolving social network models, the probability for the formation of new links in the (outer) social network between the agents as well as the probability for dispatching an existing link are affected by the similarity between the attitudes of the agents\cite{Connected09, HomophSoc}.

The novelty of our approach lies not in discussing the above-mentioned human characteristics, rather in demonstrating their fundamental effect both on our attitudes and on the structure of social networks, and, importantly, the way they can be integrated into agent-based models.  We are able to show that an initially random social network between "clones" (agents with identical internal belief system) -- depending on the actual set of parameters --  may facilitate the emergence of a rich variety of human group structures, including consensus, fragmentation and polarization. 
 In addition, we also find that the "repulsive interaction" effect (a.k.a "distancing", when agents' opinion further depart in case of a certain amount of disagreement\cite{RepulsiveEffect}) is a natural sequel of the effort of keeping the coherence level as high as possible, and as such, there is no need to introduce this rule "by hand"\cite{NoDistancing}.

\section*{Model description}

Our approach considers an evolving social network between agents that can communicate with their neighbours over the links. This network is directed, where the out-links of an agent indicate with whom it may initiate a communication act. We assume that each agent has an internal state (what we can refer to as a belief system) that affects the communication with the out-neighbours and also the establishment of new connections. Furthermore, we also assume that the communication acts affect the internal state of the recipient agents, thus, in parallel with the network structure, the internal belief systems of the agents is also changing over time. 

This internal belief system is modeled by a network (graph) in which the nodes are \textit{beliefs} or \textit{concepts} (which we use as synonyms in the present paper), node \textit{weights} -- a scalar value ranging from $-1$ to $+1$ -- represent the agents' \textit{attitudes} towards these concepts with negative values indicating negative sentiments and positive values showing support, and finally, the \textit{edges} of the graph reflect the \textit{level of association} between the given nodes (concepts). 

That is, there are two types of networks being studied in parallel in our simulations: (i) a social network, where nodes represent agents and edges indicate the likelihood of communication, and (ii) belief networks, where nodes represent concepts or beliefs that are interconnected. Each agent, naturally, possesses a unique belief system.

\subsection*{Overview of the main concepts}

\begin{figure}[ht!]
\centering
\includegraphics[width=\linewidth]{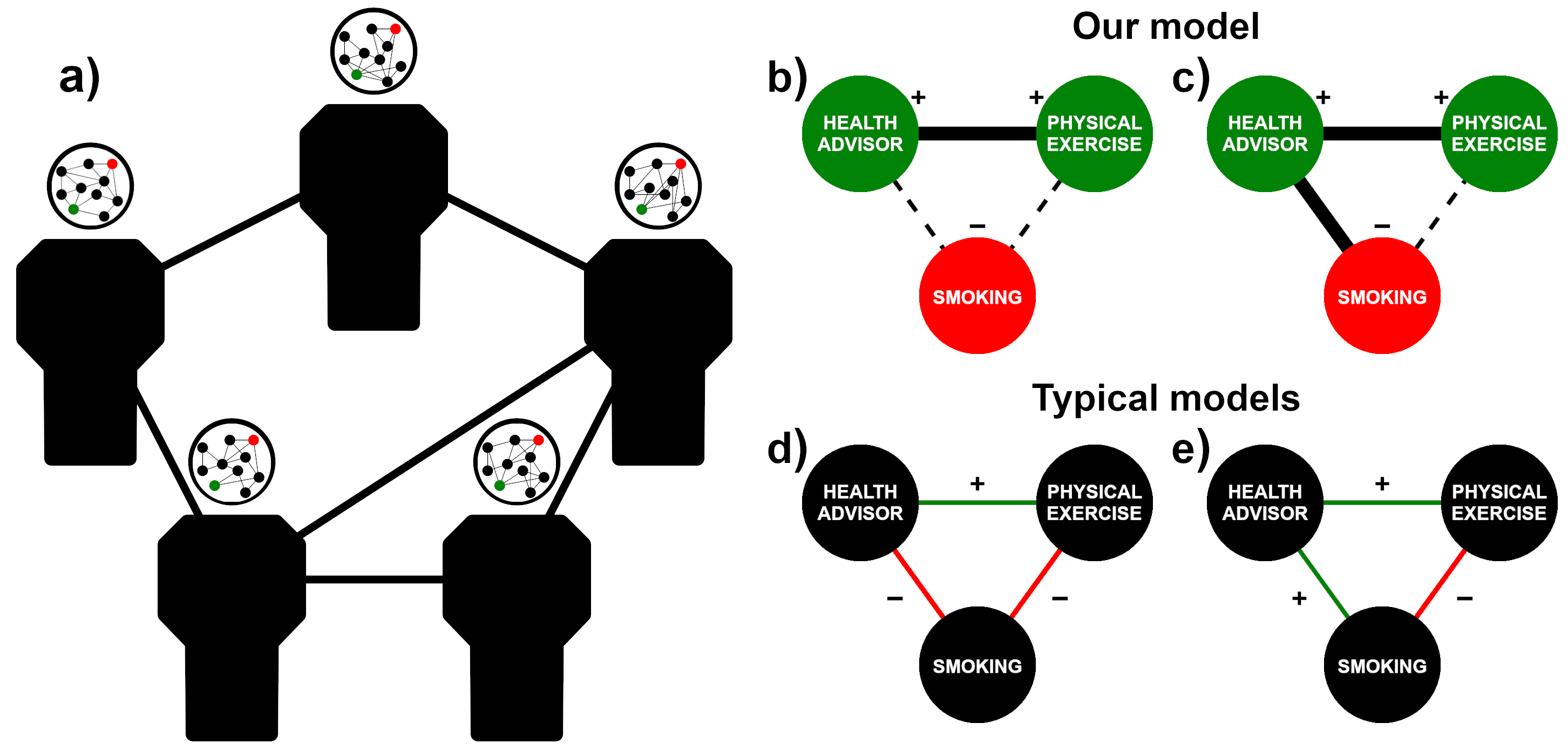}
\caption{\textbf{Social network of agents with internal belief systems.} a) Illustration of the social network of the agents where the belief systems of the agents are represented by additional internal networks. The coloured nodes correspond to the concepts with fixed attitudes, one having a constant +1 value (green) and the other having a constant -1 value (red). b) A 'coherent' triangle of concepts in our model, where two concepts with the same signed attitude (positive in this case) are connected by a strong association link while the connections between concepts with opposite signed attitudes are weak. c) An 'incoherent' triangle in our model, where a strong connection appeared between two concepts having opposite signed attitudes. d) A 'coherent' triangle of concepts in models which operate with signed connections between the concepts. e) An 'incoherent' triangle in models based on signed connections in the belief system.}
\label{fig:trianglesCombined}
\end{figure}

The most important traits we incorporate into our model are the following:
\begin{enumerate}
    \item \textit{Internal state represented by a weighted graph.} In humans, opinions and beliefs are never isolated; every idea or belief is interconnected with others. As a matter of fact, humans cannot even remember something unless it is associated with something meaningful\cite{EBGoldsteinCognPsy}. In other words, concepts and beliefs are arranged into an organized framework, often referred to as a \textit{belief system}. According to that, in our model the opinion or internal state of the agents is not a scalar or even a vector, instead it is represented by a small internal network between the beliefs or concepts, as illustrated in Fig.~\ref{fig:trianglesCombined}a. In this approach, the non-negative link weights (falling into the $[0,1]$ interval) represent the level of association between two concepts, whereas signed node-weights (falling into the $[-1,1]$ interval) indicate the agents attitude towards the given concept or belief, with negative values indicating condemnation and positive values referring to support. 
   
    \item \textit{Communication is enhancing the association between the topics.}  New information often comes in the form of connecting originally unrelated concepts, and these new associations may immediately entail the re-evaluation of the pre-existing beliefs and attitudes. In accordance with that, in our model, information, coming through social interactions change the connection weights between beliefs in the internal belief system of the interacting agents. More concretely, during a communication act, the initiator agent is choosing a pair of concepts (serving as the "topic" of the conversation) according to a probability proportional to their connecting weight in its belief system,  and as a result of the communication act, the association (link-weight) of the same link is enhanced also in the belief system of the recipient agent. However, in order to prevent all link-weights in the belief system approaching 1, after the modification of the association weight between the topics of the communication, the link-weights in the belief system of the recipient agent are re-normalised so that the sum of the weights in its belief system remains constant.
    
    \item \textit{Cognitive dissonance avoidance.} A key characteristic of human belief systems is our strong desire to maintain coherence among our beliefs, while striving to steer clear of the uncomfortable sensation known as cognitive dissonance, which is basically the feeling of \textit{in}coherence. 
    Based on that, one of the main driving forces of our model is that after communication acts, the recipient agents try to minimise the cognitive dissonance of their internal belief system by modifying their attitudes towards the concepts, that is, the node-weights in the internal state representation. 

    To illustrate how cognitive dissonance may arise in our approach, let us imagine someone having a \textit{health advisor} recommending \textit{physical exercise}, which are connected to each other by a strong edge, since they are related (associated) to each other, as shown in Fig.~\ref{fig:trianglesCombined}b. 
    In contrast, \textit{smoking} is unrelated to physical exercise (since they are activities that people typically do not pursue in parallel), so the edge weight between them is zero or close to zero. Similarly, the edge weight between the \textit{health advisor} and smoking is also zero, because they are not associated either. (Fig.~\ref{fig:trianglesCombined}b). In contrast, if our agent gains knowledge that the health advisor is a smoker, the edge between them becomes strong, as shown in Fig.~\ref{fig:trianglesCombined}c. Let us also assume that the agent has a positive attitude towards physical exercise and the health advisor (green colour) and negative attitude towards smoking (red colour) in these examples. In the case of Fig.\ref{fig:trianglesCombined}b, concepts with the same sign are connected by a strong connection, whereas concepts with opposing sings are not connected (or are only very weakly connected), thus, we have coherence. In contrast, in the case of Fig.\ref{fig:trianglesCombined}c, the strong connection between smoking and the health advisor, having opposite signed node-weights, leads to in-coherence, inducing cognitive dissonance. 

    According to the above, the \textit{coherence} of a single pair of concepts in the belief system can be defined simply as the product of the (signed) node-weights and the weight of the association link between them, whereas the coherence level of the entire belief system can be given as the sum over the pairs of concepts. 
    For comparison, in Fig.\ref{fig:trianglesCombined}d-e we show the traditional approach (which is motivated by the Balance Theory\cite{HeiderBT, ESTRADA201970}) for representing coherence and in-coherence among concepts via signed connection weights. Here, by multiplying the signs of the three edges appearing in a triad\cite{DoB,Rodriguez16, PhysRevE.90.042802} we may end up with a positive or a negative result, former indicating a stable, coherent triad and the latter corresponding to an unstable, in-coherent one.
\end{enumerate}



Beside cognitive dissonance avoidance, another fundamental feature of the human mind is that we share some universally positive concepts which remain positive independently of our education or socialization. For example, due to its deep biological and psychological roots, the concept of \textit{mother} is typically associated with unconditional love, care and protection, and as such, forms an unconditional, invariably positive concept\cite{moth1, moth2,moth3}. In contrast,  for example \emph{fear} or \emph{death} are uniformly negative concepts\cite{fear, death}. Furthermore, other concepts might also tend to be "constant" positive or negative, depending on the culture the agent belongs to\cite{CultDiff}. That is, some other constant negative or positive attitudes might differ from culture to culture. In our model, such concepts can be incorporated by nodes in the belief system that have an unalterable weight of either $+1$ or $-1$. For sake of simplicity, in the present work we have incorporated a single concept with constant positive weight of $1$ and another single concept with constant negative weight of $-1$. These two concepts are the same for all the agents, depicted by the green and the red colour in Fig.~\ref{fig:trianglesCombined}a.

In our approach the time evolution of the social network is manifested in the change of the connection weights between the agents, where each link-weight can take a value in $[0,1]$. After a communication act initiated by agent $i$ with an other agent $j$, the weight of the out-link on $i$ towards $j$, denoted by $w_{ij}$ is increased or decreased based on the similarity between the attitudes of the two agents with regard to the topic of the communication (a pair of concepts). In parallel, the weight of the out-link on the recipient node $j$ towards $i$ is also updated, however here the cognitive dissonance $j$ has possibly experienced during communication is also taken into account. Naturally, since cognitive dissonance leads to discomfort, if the coherence level of $j$ is lowered due to the communication act, this will have a negative effect on $w_{ji}$.

\subsection*{Detailed model description}

Our model features $N$ agents, each with their own belief system containing $M$ nodes (concepts). These concepts might range from political positions to football teams and public figures, among others. Each agent has an attitude towards these concepts, represented by a node-weight which is quantified on a scale from $-1$ to $+1$, indicating completely negative to completely positive views, respectively. The un-directed connections between concepts are also weighted with link-weights ranging from $0$ to $1$. For instance, if an agent learns that their favorite football team is implicated in a match-fixing scandal, the weight on the edge connecting these two concepts would increase.


Based on the node-weights, two concepts can either reinforce each other (if their sign is the same), leading to \textit{reassurance}, or conflict with one another (one being positive, the other negative), causing \textit{cognitive dissonance}. For instance, if a politician whom a person favors supports a policy that the individual also endorses, linking the politician with the policy causes reassurance (it increases the level of coherence). Conversely, if that same politician is associated with a disfavoured political agenda, then this association cases cognitive dissonance. We assume that reducing cognitive dissonance is more crucial than merely achieving reassurance (tuned by the parameter $d$)\cite{CognDissA}. Accordingly, for a  pair of concepts $\alpha$ and $\beta$ with respective node-weights (attitudes) $a_{\alpha}$ and $a_{\beta}$, the impact on the coherence level of the agent is defined as
\begin{equation} 
        C_{\alpha\beta} = \begin{cases}
                d \cdot a_{\alpha} \cdot a_{\beta} \cdot B_{\alpha\beta}, & \text{if } a_{\alpha} \cdot a_{\beta} < 0 \\
                a_{\alpha} \cdot a_{\beta} \cdot B_{\alpha\beta}, & \text{otherwise},
            \end{cases}
            \label{eq:coherence_single}
\end{equation}
where $d$ is the dissonance penalty parameter and $B_{\alpha\beta}$ indicates the link-weight between $\alpha$ and $\beta$ in the belief system. Although in real-life societies $d$ probably follows a distribution, in the present model -- for sake of simplicity -- we handle it as a constant parameter.

The total coherence level of the agent can be calculated by summing over all concept pairs as
\begin{equation} 
    C_{\rm tot} = \frac{2}{M (M-1)}\sum_{\alpha=1}^{M}\sum_{\beta=\alpha+1}^M C_{\alpha\beta},
    \label{eq:coherence_tot}
\end{equation}
where we have divided the sum by the total number of concept pairs in the belief system. In this way, a fully coherent belief system has a coherence of $C_{\rm tot}=1$. As we shall detail later, the agents try to maximize the coherence level of their belief system after each communication act in which they took part as a 'receiver'. 


The $N$ agents are all part of a directed and weighted social network, where link weights can range between 0 and 1. The outgoing links of the agents represent the willingness that they choose to communicate with the other agents. The social network evolves through this communication between individuals. 

At the beginning of each iteration we choose uniformly at random the agent who will be the initiator of the communication act. 
The probability that the chosen initiator agent $i$  will communicate with an other agent $j$ is depending mainly on the strength of the link pointing from $i$ to $j$, denoted by $w_{ij}$. Furthermore, we also incorporate {\it triadic closure} \cite{Watts1998, triadic_closure} into the model, corresponding to a widely used concept in 
network science, referring to an increased likelihood for the formation of triangles, resulting from connecting to a friend of a friend. 
With formula, the probability for communicating with $j$ is proportional to
\begin{equation} 
    \hat{p}_{ij} = w_{ij} + \frac{\sum_{q\neq i} w_{iq}  w_{qj}}{TCA(\sum_q w_{iq})}, 
    \label{eq:comm_partner_prob}
\end{equation}
where  $TCA$ is a (monotonously increasing) \textit{triadic closure affinity} function depending on the out-strength (sum of the out-weights) of $i$ (see Eqs.~\ref{TCA_x}, \ref{TCA_x_squared} and \ref{TCA_4th_power}). The role of $TCA$ is to ensure that triadic closure has a stronger effect on the communication of agents with a low strength value and a weaker effect on 'hubs' that have a high number of strong out-links. The actual probability for choosing $j$ as the recipient in the communication act is simply given by normalising the $\hat{p}_{ij}$ values as
\begin{equation} 
    p_{ij} = \frac{\hat{p}_{ij}}{\sum_j \hat{p}_{ij}}. 
\end{equation}

Once the recipient agent is fixed, a pair of concepts $\alpha$ and $\beta$ are chosen to serve as the 'topics' of the communication. The probability for choosing $(\alpha,\beta)$ is proportional to the weight $B_{\alpha\beta}$ between $\alpha$ and $\beta$ in the belief system of the initiator agent $i$. As the result of the communication, the weight of the connection between the same two concepts is increased by a random number between 0 and 1 in the belief system of the recipient agent $j$. (Naturally the new $B_{\alpha\beta}$ in agent $j$'s belief system is capped at 1 if an overshoot should occur). After that, all weights in the belief system  of $j$ are normalised to keep the total sum of edge weights constant through the process. 



To account for the desire to maintain a coherent belief system, the recipient agent $j$ is also permitted to adjust the node-weights within its belief system in response to the updated $B_{\alpha\beta}$ values. For instance, if agent $j$ seeks to reduce cognitive dissonance between their favorite football team and the concept of match-fixing, they might choose to either diminish their positive attitude towards the team or enhance their acceptance of match-fixing. In our model, this is captured through a stabilization process, where the agent may attempt to randomly alter its node-weights over $b$ iterations. During each iteration, an initial push value is generated within the range of $-1$ to $1$. If the sign of this push aligns with that of the original attitude value, the push is scaled by the distance from the corresponding extreme pole before being applied to the attitude, otherwise it's default value is used.
This way, the attitude value will never go below $-1$ or above $1$. 

Following this adjustment, the agent recalculates the overall coherence level using the updated attitude value based on equation (\ref{eq:coherence_tot}). The agent accepts the new attitude value according to a formula that is widely-used in optimization problems, namely that if the condition $\exp(\frac{k\cdot C_{\rm diff}}{T})>r$ is met, where $k$ represents the iteration count, $C_{\rm diff}$ is the difference between the new and old coherence levels, $T$ is a temperature-like parameter, and $r$ is a randomly generated number uniformly distributed between $0$ and $1$.


The recipient agent performs the above stabilization process, first on each of the two concepts discussed during the communication for $b$ iterations, then on randomly chosen concepts for another $b$ iterations. In total, the number of stabilization processes performed is $3b$.
A flowchart of the stabilization process can be seen in the supplementary material.

We note that the above rules for the update of the association weights between the concepts and the attitudes at communication acts inherently modify the belief system even when a pair of identical agents in full agreement with each other interacts. For example, it is possible that the increase of the association weight between the 'topics' of the communication together with renormalisation of the aggregated association weights (slightly reducing all the other weights) may decrease the coherence for the recipient agent. As a result, this agent may change its attitudes towards some of the concepts, introducing disagreement between the initially identical agents.

Naturally, the communication also affects the links in the social network. For the initiator node $i$, this is based only on the attitudes of the beliefs discussed during the communication. Therefore, the update rule for the weight of the out-link from $i$ to $j$ is defined as
\begin{equation} 
    w_{ij}(t) = w_{ij}(t-1) + a_{\alpha}^{(i)} \cdot a_{\alpha}^{(j)} + a_{\beta}^{(i)} \cdot a_{\beta}^{(j)}, 
    \label{eq:soc_change_init}
\end{equation}
where $a_{\alpha}^{(i)}$ denotes the attitude of the $i$ regarding concept $\alpha$, etc. In parallel, the recipient agent $j$ will also change the link-weight of its out-link pointing towards the initiator agent $i$, but with the new attitudes after the stabilization process, given as
\begin{equation} 
    w_{ji}(t) = w_{ji}(t-1) + a_{\alpha}^{(i)} \cdot a_{\alpha}^{(j) {\rm new}} + a_{\beta}^{(i)} \cdot a_{\beta}^{(j) {\rm new}}. 
    \label{eq:soc_change_reciev}
\end{equation}
In the case of Eqs.(\ref{eq:soc_change_init}-\ref{eq:soc_change_reciev}), if the new weights would fall outside of the $[0,1]$ interval, they are set to $0$ or $1$, respectively. 

The two most important parameters in our simulations were:

(1) The dissonance penalty, $d$, is the parameter appearing in (\ref{eq:coherence_single}) that controls the strength with which an agent can or cannot tolerate cognitive dissonance, with higher values referring to smaller tolerance. In other words, it reflects the strength by which an agent is “bothered” by experiencing cognitive dissonance.

(2) The triadic closure affinity, $TCA$, is a function that appears in (\ref{eq:comm_partner_prob}), controlling the likelihood with which agents choose to communicate with a 'friend of a friend' in the social network, instead of their own neighbours. 
We study 3 cases:
        \begin{equation}
            TCA(x) = x
            \label{TCA_x}
        \end{equation} - This is the least strict case, that is, the most open for communication with friends of friends. 
        
    \begin{equation}
        TCA(x) = x^2
        \label{TCA_x_squared}
    \end{equation} - This is a stricter case, where the number of friends and the normalizing factors do not scale linearly. This means that agents communicate less through their friends' social networks if they have enough people in their own circles.
    \begin{equation}
        TCA(x) = x^4 
        \label{TCA_4th_power}
    \end{equation}
    - This is the strictest case, where the number of friends reduces the willingness to communicate outside of direct friends even more.

\section*{Simulation results}

Our model consisted of 100 agents in a fully connected social network with all link-weights being equal to $w_{ij}=1$ at the beginning of each trial. The agents were all clones (identical copies) of an initially generated agent. The belief system size was $M=10$, having one concept with constant positive attitude $a_{\alpha}=1$ and also another concept with constant negative attitude $a_{\beta}=-1$. The temperature parameter was set to $T = 0.01$. 

For each trial, we set the parameters and ran the simulation until it converged and reached its final state, where the main structural properties of the social network were no longer changing. The precise criteria of convergence are described in the Methods. We note however that links are constantly formed and deleted in our model, thus, these end states were static only in the statistical sense. The results below refer to these steady, final states. 




\subsection*{Polarisation and fragmentation of the social network}

\begin{figure}[h!]
\centering
\includegraphics[width=\linewidth]{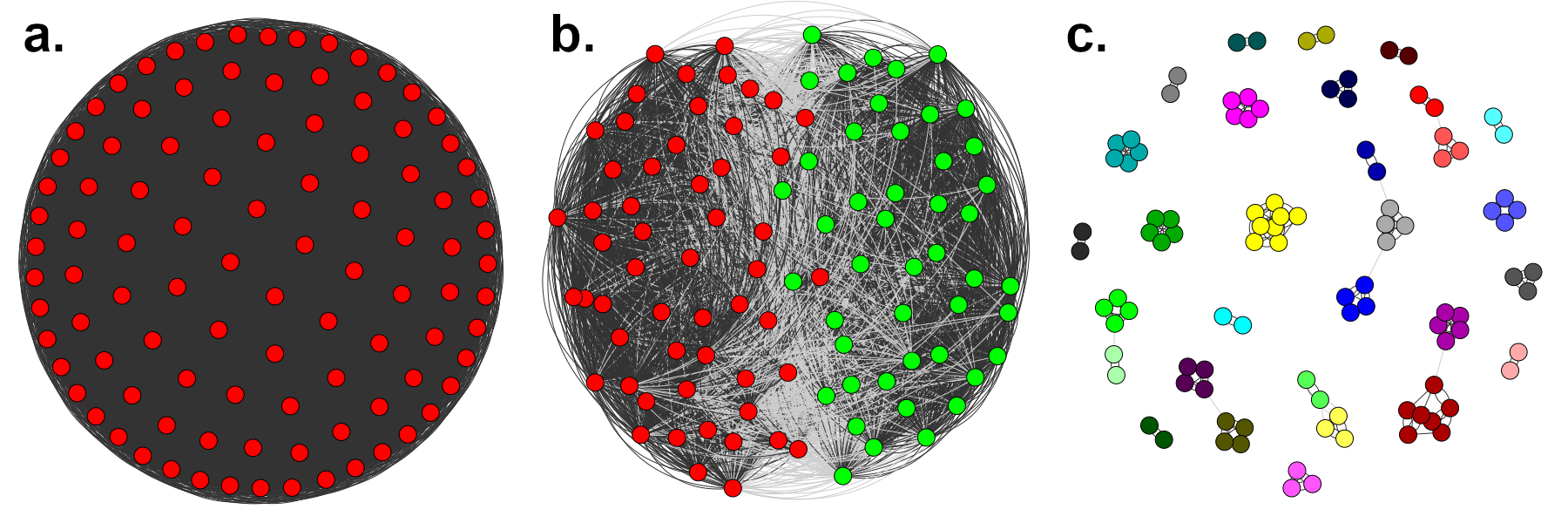}
\caption{\textbf{Stable end-states of the social network.} Three different states of the social network after convergence in three simulations with different parameter settings. \textbf{a)} With $d=2$ and $TCA=x$ consensus emerges. \textbf{b)} $d=4$ and $TCA=x$ results polarisation, where two communities form, indicated by colours (extracted using the Leiden algorithm~\cite{Leiden}). \textbf{c)} When using $d=4$ and $TCA=x^4$, we can observe fragmentation of the network into small isolated components.}
\label{fig:soc_end_states}
\end{figure}

Our general observation is that the defined social network model can reach a wide variety of stable end states, including consensus (where basically all agents have a more or less similar opinion and are willing to communicate with each other), polarisation (where two large groups are formed with a clear opinion difference) and fragmentation (where the network falls apart into small isolated components). We note that although the social network always starts from a fully connected network with unit link weights, during the time evolution of the system whenever a link weight $w_{ij}$ becomes zero, that link can be considered to be removed from the network. As an illustration of the wide range of possible behaviours of our model, in Fig.\ref{fig:soc_end_states} we show the layouts of three networks, corresponding to the stable end states of simulations at different parameter settings.


The final states are shaped by the dissonance penalty $d$ and the $TCA$ function. With a low dissonance penalty (referring to high tolerance towards cognitive dissonance), agents are less likely to alter their attitudes, resulting in no change within the social network since the agents are identical to each other (Fig.\ref{fig:soc_end_states}a). However, when the dissonance penalty is higher, agents may deviate from their initial set of attitudes.



Regarding the $TCA$ parameter, which determines the probability of interacting with friends of friends according to Eqs.(\ref{TCA_x}-\ref{TCA_4th_power}), we observed that when using a simple linear function $f(x) = x$, at higher values of the dissonance parameter $d$ the social network generally divides into two communities (Fig.\ref{fig:soc_end_states}b). These correspond to dense subgraphs that are more loosely connected to each other. Communities, also called as modules, groups or clusters are known to occur not only in social networks but also in various other types of networks, and community finding is in general an intensively researched topic in network science~\cite{Fortunato_coms,Fortunato_Hric_coms,Cherifi_coms}. In the present work, we used the Leiden algorithm~\cite{Leiden} to identify communities, a highly efficient method known for producing high-quality partitions of input networks based on modularity~\cite{Newman-Girvan_modularity} (corresponding to the most widely used metric for evaluating community quality). Both the Leiden algorithm and the modularity are briefly described in Methods.

When nonlinear $TCA$ functions such as $f(x) = x^2$ are applied, the average number of groups increases, typically ranging between 3 and 4. If the nonlinear effect is made even stronger with a function like $f(x) = x^4$, we can see the groups fragmenting further, as agents become increasingly unlikely to expand their social circle once they have established their own friends. In such cases, the social network in the end state is composed of numerous small isolated components (Fig.~\ref{fig:soc_end_states}c).


\subsection*{Divergence of the attitudes}  

\begin{figure}[h!]
\centering
\includegraphics[width=\linewidth]{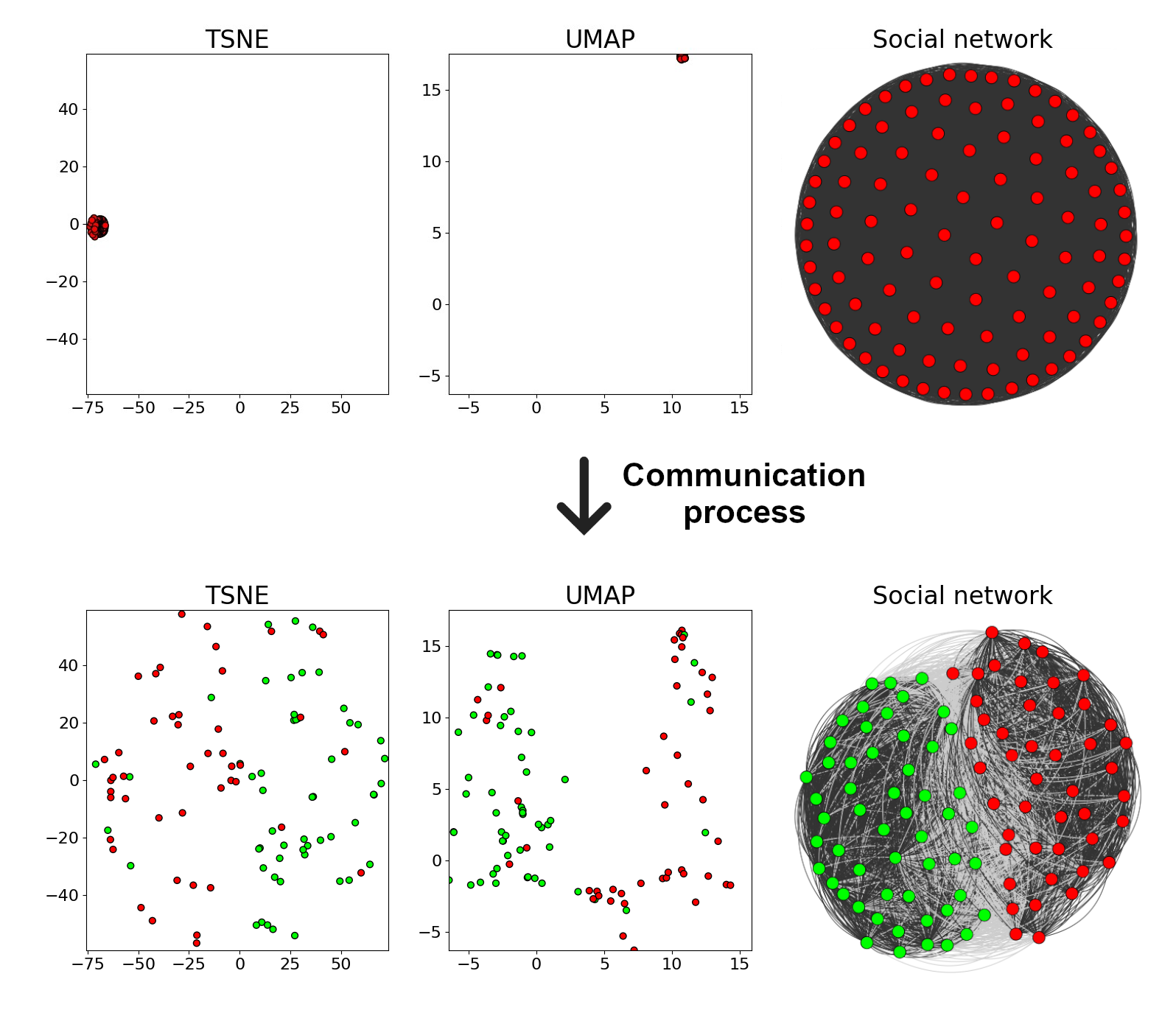}
\caption{\textbf{Attitude embeddings and social network after communication process.} The top row shows the embedding of the attitudes (node-weights in the belief systems) according to the TSNE (left) and UMAP methods with the social network displayed on the right for the initial state of the simulation. Here the social network is fully connected and is consisting of agents with identical belief systems. At this state, the clones form a dense cluster in the attitude space, with only minor differences due to the stochastic nature of the TSNE/UMAP embeddings. In the bottom row we show similar results after a certain amount of communication, where the social network forms two densely connected communities (marked by the different colours). These communities can also be seen in both embeddings, roughly corresponding to a left/right split.}
\label{fig:embedding_change}
\end{figure}

In all simulations, our agents were initially identical clones, sharing the same belief systems and attitude values. However, as the interaction progressed, usually their belief systems began to diverge from the initial state. Especially at larger $d$ values (corresponding to lower tolerance levels towards cognitive dissonance), at some point during the simulation, the occurrence of a drastic change was typical during which the agents divided into two (or more) communities  (groups, clusters, etc.) with respect to both their belief systems and the structure of the social network. This is illustrated in Fig.\ref{fig:embedding_change}, comparing the initial state of the system with a later stage after a certain amount of communication. 


To visualize the similarities or differences between the belief systems of the agents, we employed $t$-distributed Stochastic Neighbor Embedding\cite{tsne} (TSNE), as shown in the left column, and Uniform Manifold Approximation and Projection\cite{umap} (UMAP), displayed in the middle column. Both methods use the attitudes (node-weights) of the agents as input vectors and provide a nonlinear projection of this data into two dimensions. (A brief description of these embedding methods is provided in the Methods section). As illustrated in the top row of Fig.~\ref{fig:embedding_change}, at the initial state where the agents' belief systems are identical, they form very tight clusters in the attitude embedding spaces, with minimal variance occurring solely due to the stochastic nature of the embedding algorithms.

In contrast, after certain amount of communication, the cloud of agents splits into two clusters according to both TSNE and UMAP, as indicated in the bottom row if Fig.\ref{fig:embedding_change}. Simultaneously, the structure of the social network has also transformed (right column of Fig.\ref{fig:embedding_change}), evolving from an initially fully connected graph into a network comprising two dense communities. These communities (also called as modules or groups) were located with the Leiden algorithm\cite{Leiden}. (A brief description of the Leiden approach is given in Methods). The colouring of the agents in the left and middle panels reflect their community membership in the social network. As it can be seen, there are agents who are on one side of the attitude space but still belong to the opposing social group. These are those agents who have already changed their attitudes but haven't communicated enough yet to leave their current community.

In order to quantify the similarities and differences between the attitudes of the agents at the group level, we define a quantity we call \textit{attitude homogeneity}, $AH\in[0,1]$, calculated from pairwise comparisons of agents' attitudes and relying on the absolute difference between the node-weights. (The precise definition of $AH$ is given in Methods). This metric takes a value of $1$ if, and only if all considered agents have exactly the same attitudes. In parallel, we also define \textit{extremism}, $E$ corresponding to the average of the absolute magnitude of the node-weights in the belief system. (A larger $E$ values indicates that the given agent has more attitudes falling closer to the extreme $+1$ or $-1$ values). 

\begin{figure}[h!]
\centering
\includegraphics[width=\linewidth]{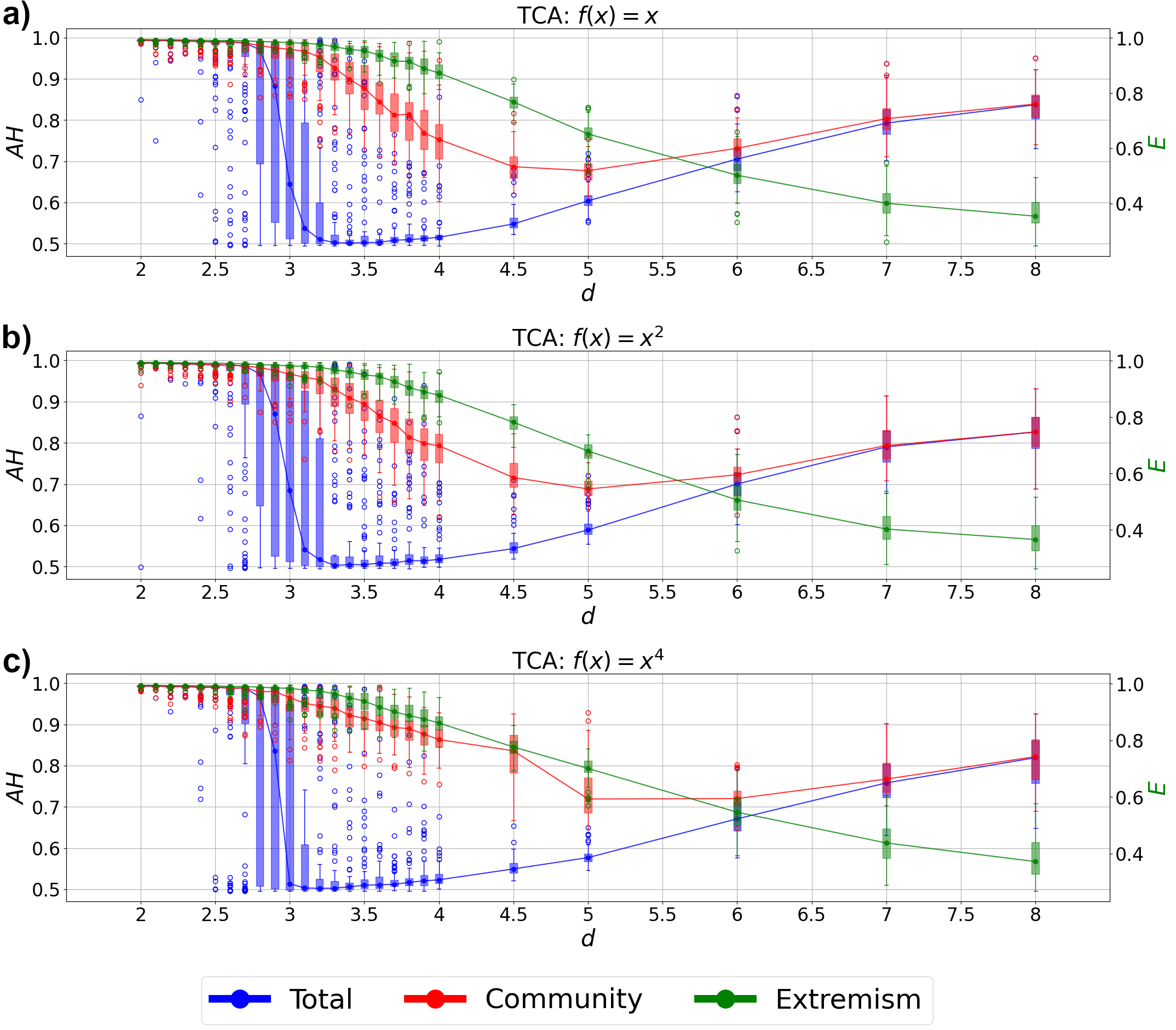}
\caption{\textbf{Attitude homogeneity (AH) and extremism (E) values.} Boxplot of the $AH$ and $E$ over 100 runs as a function of the dissonance penalty $d$. The $AH$ calculated for the entire network is shown in blue, whereas the $AH$ evaluated inside communities (found by the Leiden algorithm) is shown in red. In parallel, the extremism is also plotted in green using the right vertical scale. The median values are connected by continuous curves. The three panels correspond to different $TCA$ functions as indicated by the panel titles. The total homogeneity values begin to diverge when approaching $d=3$, while the community homogeneity remains high. As extremism starts to drop, around $d=6$ the majority of agents form a moderate core, which decreases the differences between attitudes, and the total and community homogeneities become similar again.}

\label{fig:combined_attitude_homogeneity}
\end{figure}

In Fig.\ref{fig:combined_attitude_homogeneity} we show the attitude homogeneity $AH$ and extremism $E$ at the end state of the simulations as functions of the dissonance penalty $d$ for the three different $TCA$ functions.  
According to that, the attitude homogeneity of the entire network (shown in blue) drops sharply around a dissonance penalty value of $d=3$. In other words, above a certain level of \textit{in}tolerance towards cognitive dissonance, attitudes begin to rapidly diverge. As we shall show later, this is also the point where we can observe an important change in the network structure of the end states as well, where for low $d$ values we usually observe one single community and for $d>3$ two or more communities emerge. Therefore, when multiple communities form in the social network, members belonging to the same communities develop similar attitudes that are at the same time different from the attitudes of the agents in other communities. Accordingly, the homogeneity of the communities (shown in red) lacks this sharp drop and shows a slower decrease as a function of $d$.

The intuitive explanation for the above behavior derives from the avoidance of cognitive dissonance, which acts as a fundamental drive in our model. Typically, agents experience the most coherent state when their attitudes are close to one of the two extreme values, -1 or 1. Nonetheless, as the penalty for cognitive dissonance ($d$) increases, a new optimal point emerges around 0, marking an "indifferent" stance by which the agent avoids cognitive dissonance (as well as reassurance). This state is generally temporary, with agents often reverting to one of the extreme attitudes, possibly even switching to the opposite compared to their initial attitudes. Consequently, repeating the interaction process may lead the group to adopt a full spectrum of attitudes, with an equal distribution of agents at both extremes.

However, if we increase $d$ further, we can reach a point where this "neutral" stance becomes stable, in which case the attitudes of all agents will stay more similar overall. This represents the formation of a "moderate" majority, who are neutral on almost all issues. They are usually flanked by a smaller number of extremists on both sides. 

The above explanation is also supported by the behavior of the extremism values (shown in green), which are monotonically decreasing as a function of $d$ in Fig.\ref{fig:combined_attitude_homogeneity}. This indicates that the increase of homogeneity in the large $d$ regime is accompanied by more and more agents taking up "neutral" or less extreme attitudes.



\subsection*{Structural changes in the social network}

\begin{figure}[h!]
\centering
\includegraphics[width=\linewidth]{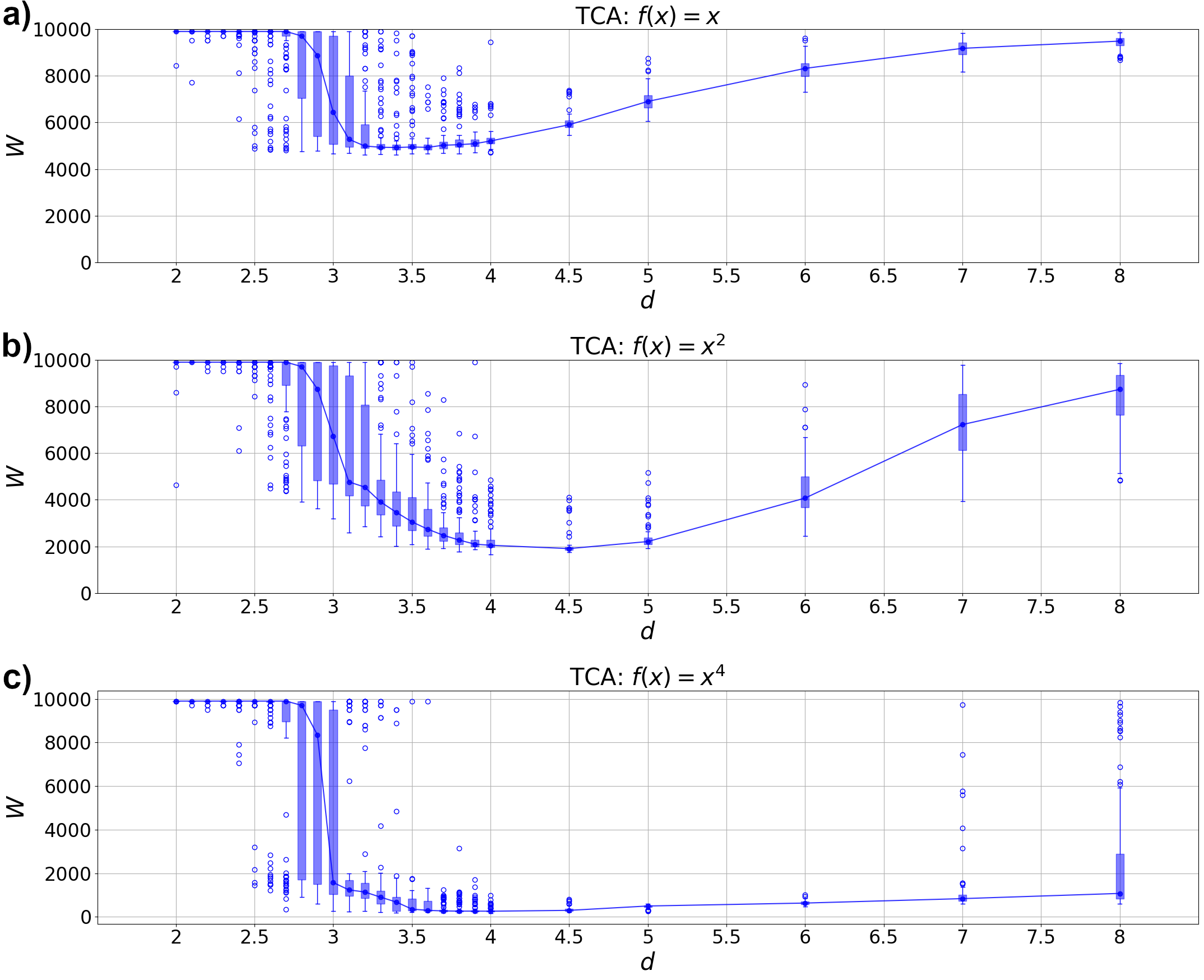}
\caption{\textbf{Sum of the link weights in the social network.} We show the aggregated link weights, given by $W=\sum_{ij}w_{ij}$ after convergence, as a function of the dissonance penalty $d$. The symbols represent a standard box plot over 100 runs for each parameter setting according to $d$, and the median values are connected by a continuous curve. The three panels correspond to different $TCA$ functions as indicated in the panel titles. Close to $d=3$, $W$ starts to decrease rapidly in all cases, indicating the disappearance of social links in the network. However, depending on the $TCA$ function, the remaining number of edges can vary from around half of all edges to a few hundred. For higher $d$ values, this effect decreases, and the end states remain more similar to the original fully connected network.}
\label{fig:combined_edge_weights}
\end{figure}

In parallel with the changes in the belief systems of the agents, the social network can also undergo major structural reorganisation in our model. In Fig.\ref{fig:combined_edge_weights} we show the total sum of the link-weights, $W=\sum_{ij}w_{ij}$ in the end state of the simulations as a function of the dissonance penalty, $d$ for different $TCA$ functions. According to that, at low $d$ values, roughly up to $d\simeq 3$, the social network seems to retain its fully connected nature, where the sum of the link weights remain close to the possible maximal value. However, around $d\simeq 3$, for all considered $TCA$ functions a sudden drop occurs in the sum of the link weights, indicating a drastic change in the network structure. According to Fig.\ref{fig:combined_edge_weights}, for the $f(x) = x$ case around half of the possible edges remain after this drop, but for the two stricter $TCA$-s, a vast majority of the edges disappear, resulting in a sparse social network. 
In the large $d$ regime an increase in the aggregated link-weights can be observed for all studied $TCA$ functions. However, for the most strict $f(x) = x^4$ case (Fig.\ref{fig:combined_edge_weights} bottom panel) this increase is only very mild, whereas for the other two $TCA$ functions (Fig.\ref{fig:combined_edge_weights} top and middle panels) the sum of the aggregated link weights can grow back close to its initial value when $d$ becomes very high.

\begin{figure}[h!]
\centering
\includegraphics[width=\linewidth]{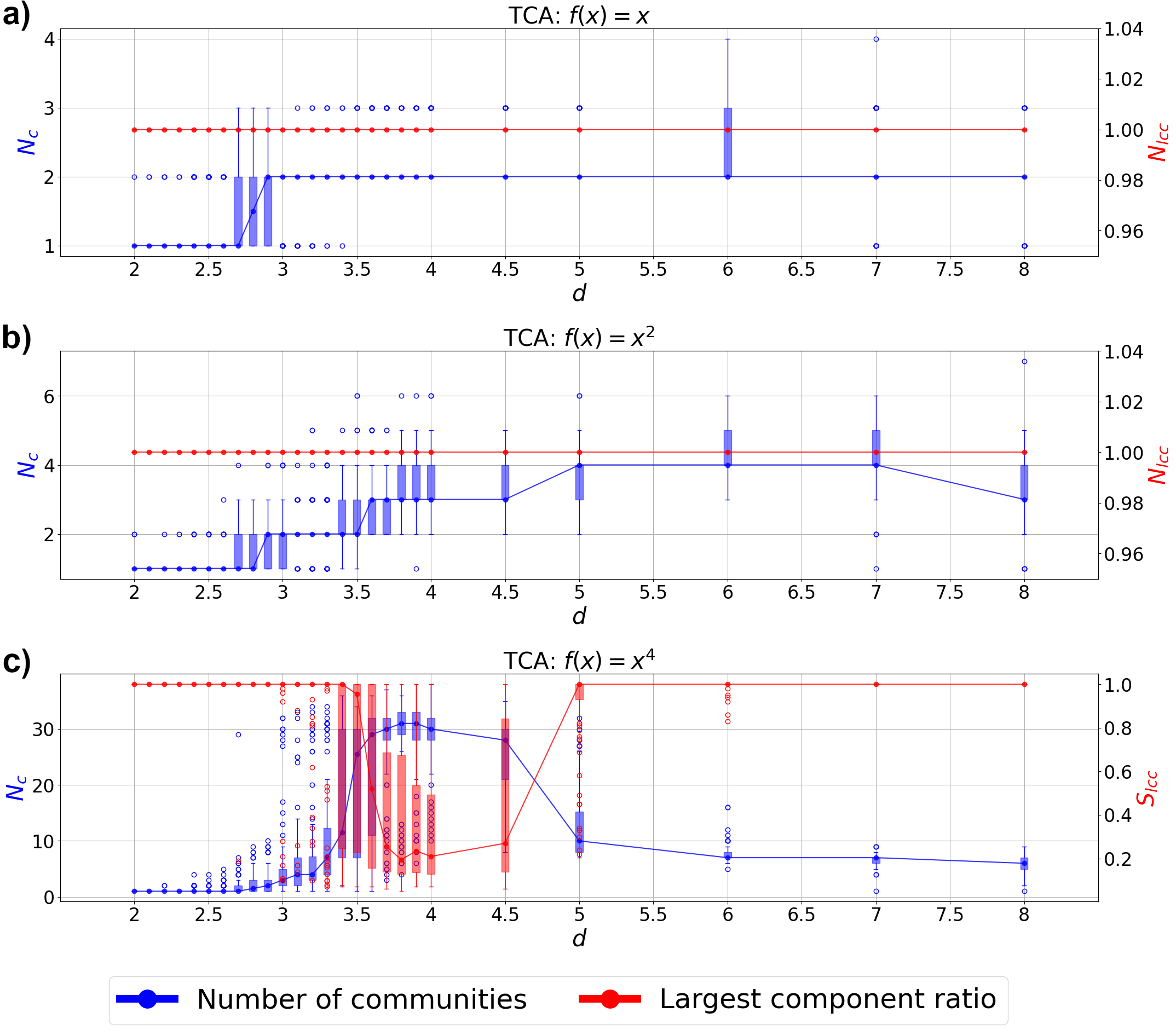}
\caption{\textbf{The average number of communities $N_c$ and the relative size of the largest connected component $S_{lcc}=N_{lcc}/N$.} The number of communities detected by the Leiden algorithm is shown on the left vertical axis as a function of the dissonance parameter, $d$. The relative size of the largest connected component compared to the total number of agents is shown on the right vertical axis. The symbols represent a standard box plot over 100 runs for each parameter setting according to $d$, and the median values are connected by continuous curves. The applied $TCA$ function is indicated in the panel titles. In all three cases, the number of communities starts to increase at $d=3$, and reaches it's maxima around $d=4$. For TCA: $f(x)=x^4$ case, this also results in the breakup of the social network into multiple components. At $d=5$, this effect starts to disappear.}
\label{fig:combined_community_stats}
\end{figure}

The drop in aggregated link weights is accompanied by the emergence of communities in the network structure. As mentioned earlier, in the present study we used the Leiden algorithm for community detection. In Fig.\ref{fig:combined_community_stats} we display the average number of communities found by the Leiden method in the end state of the simulations as a function of $d$ (blue symbols and curves). For all studied $TCA$ functions, at low $d$ values (i.e., below roughly $d\simeq 3$) the social network always consisted of a single community. This changes at the indicated transition point, above which for the $f(x)=x$ $TCA$ function (Fig.\ref{fig:combined_community_stats} top panel), in most cases the network splits into two communities, marking polarization. For the more strict $TCA$ function of $f(x)=x^2$ (Fig.\ref{fig:combined_community_stats} middle panel) the average number of communities is increasing up to roughly 4-5 communities in the larger $d$ regime. In contrast, for the most strict $TCA$ function of $f(x)=x^4$ (Fig.\ref{fig:combined_community_stats} bottom panel) the social network breaks into a larger number of small communities, consisting of only 2-6 agents each for moderate $d$ values. 

In Fig.\ref{fig:combined_community_stats} we also show the relative size of the largest connected component (red symbols and curve). 
Based on that, for the $TCA$ functions of $f(x)=x$ and $f(x)=x^2$ (Fig.\ref{fig:combined_community_stats} top and middle panels) the social network always consists of a single component, meaning that there is still communication between different communities. In contrast, for the $TCA$ function $f(x)=x^4$ (Fig.\ref{fig:combined_community_stats} bottom panel) the communities may also completely detach from the rest of the system, as indicated by the decrease of the relative size of the largest connected component roughly around $d=3.5$. In other words, for this choice of the $TCA$ function at moderate $d$ parameter values we can observe the fragmentation of the network into small isolated components that do not communicate with each other. However, when $d$ is increased to high values (that is, roughly above $d=6$), the relative size of the largest connected component increases back to one, signaling the reassembly of the network into a single connected component. This is accompanied by the decrease in the average number of communities, indicating that the formerly small isolated modules are instead organised into larger communities. This is because at higher dissonance penalties, the extremism of attitudes decreases as shown in Fig.\ref{fig:combined_attitude_homogeneity}. The outcome is an increase in agents with neutral attitudes, reducing the incentive to destroy social ties. Consequently, the social network remains more densely connected.

\subsection*{Changes in the structure of the belief systems}
 
\begin{figure}[h!]
\centering
\includegraphics[width=\linewidth]{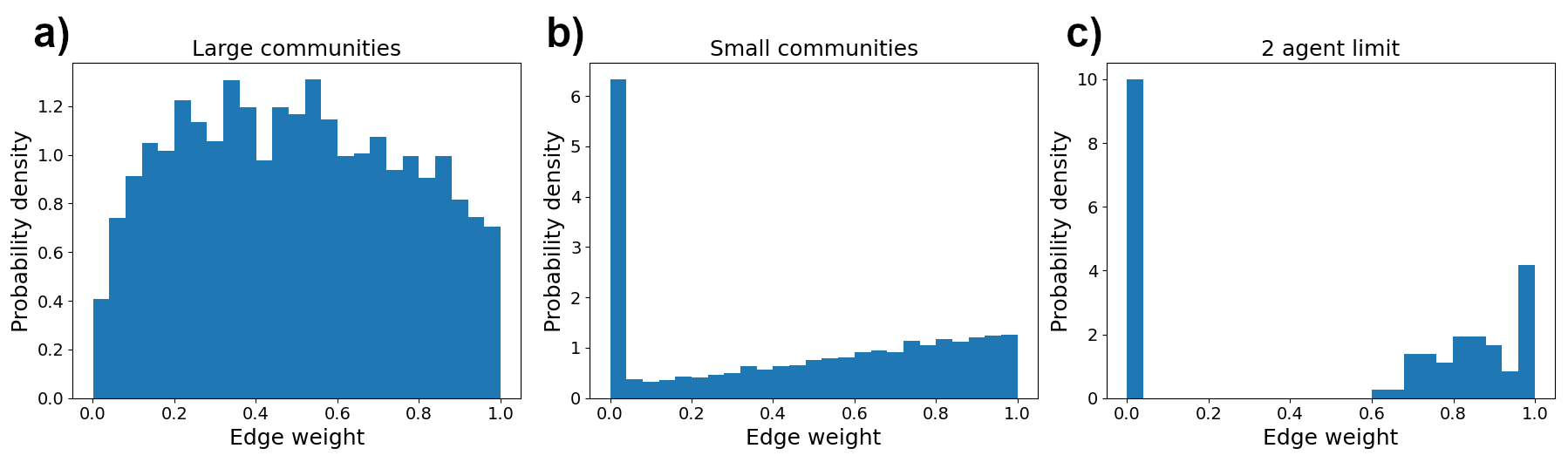}
\caption{\textbf{Distribution of the link weights in the belief system of the agents.} Panel \textbf{a)} refer to the results for parameters $d=4$ and $TCA: f(x)=x$ where the end state of the social network is composed of two large communities. Panel \textbf{b)} displays the link weight distribution of the belief systems at $d=4$ and $TCA: f(x)=x^4$, where the communities typically have only 3 to 6 members, and are completely or almost isolated from the rest of the network. Finally, panel \textbf{c)} presents the results for special simulations where only a single pair of agents were communicating at $d=4$ and $TCA: f(x)=x^4$. In the case of two agents and small communities, we can see that a large number of edges disappear from the belief system. This means that the amount of topics agents will discuss during communication becomes smaller as well.}
\label{fig:bs_end_states}
\end{figure}

 The formation of communities in the social network entangled with the development of clusters in the attitude embedding space raises the question whether we can observe related structural changes also in the belief systems of the agents. In order to investigate that, in Fig.\ref{fig:bs_end_states} we show the distribution of the link weights in the belief systems (referring to the strength of associations between concepts and beliefs) averaged over all the agents in the end-state of the simulations and also averaged over 100 runs, at three different parameter settings.   

In the case of Fig.\ref{fig:bs_end_states}a the parameters ($d=4$ and $TCA:x$) were set such that the end state of the social network showed polarisation, marked by two large communities. This meant that each agent was communicating  with roughly 50 other agents, resulting in a large variety of topics being discussed. Therefore, the resulting distribution for the connection weights between the concepts lacks any large peaks and is not far from being homogeneous. In contrast, when the communities are small and the agents communicate only with a limited number of other agents, the shape of the link weight distribution drastically changes, as indicated by Fig.\ref{fig:bs_end_states}b. Here the parameters ($d=4$ and $TCA:x^4$) were set such that the social network  fragmented into a large number of smaller, densely connected communities. Due to the limited number of communication partners, under this setting a large peak forms in the link weight distribution around zero. This represents the topics (concept pairs) that come up only with minimal probability during communication. This formation is analogous to the emergence of "interest groups" in real social systems, which are smaller communities centered around specific or specialized topics.

In order to gain an intuition about the shape of the distribution under extreme conditions we have also run simulations involving only a single pair of agents communicating with each other. Fig.\ref{fig:bs_end_states}c shows the link weight distribution of their belief systems at the end state. As it can be seen, reducing the number of communication partners to an extreme leads to a more uneven distribution, with weights either exceeding 0.6 or nearing 0. In this case, the number of topics which were involved in the communication was even more limited.

Beside the shape of the link weight distribution, we have examined the similarity between the internal structure of the belief systems between different agents as well, with a special focus on agents sharing the same community. Along this line, we introduce the \textit{belief system homogeneity}, $BH$, a quantity based on the comparison of the link-weights $B_{\alpha\beta}$ between the agents. For detailed definition see the Methods section. The belief system homogeneity calculated for a group of agents takes a value of $BH=1$ if and only if all the link-weights are identical among the examined agents (e.g., like in the initial state consisting of identical agents), while lower values indicate diversity among the connection weights (referring to the association levels) in the different belief systems.

\begin{figure}[h!]
\centering
\includegraphics[width=\linewidth]{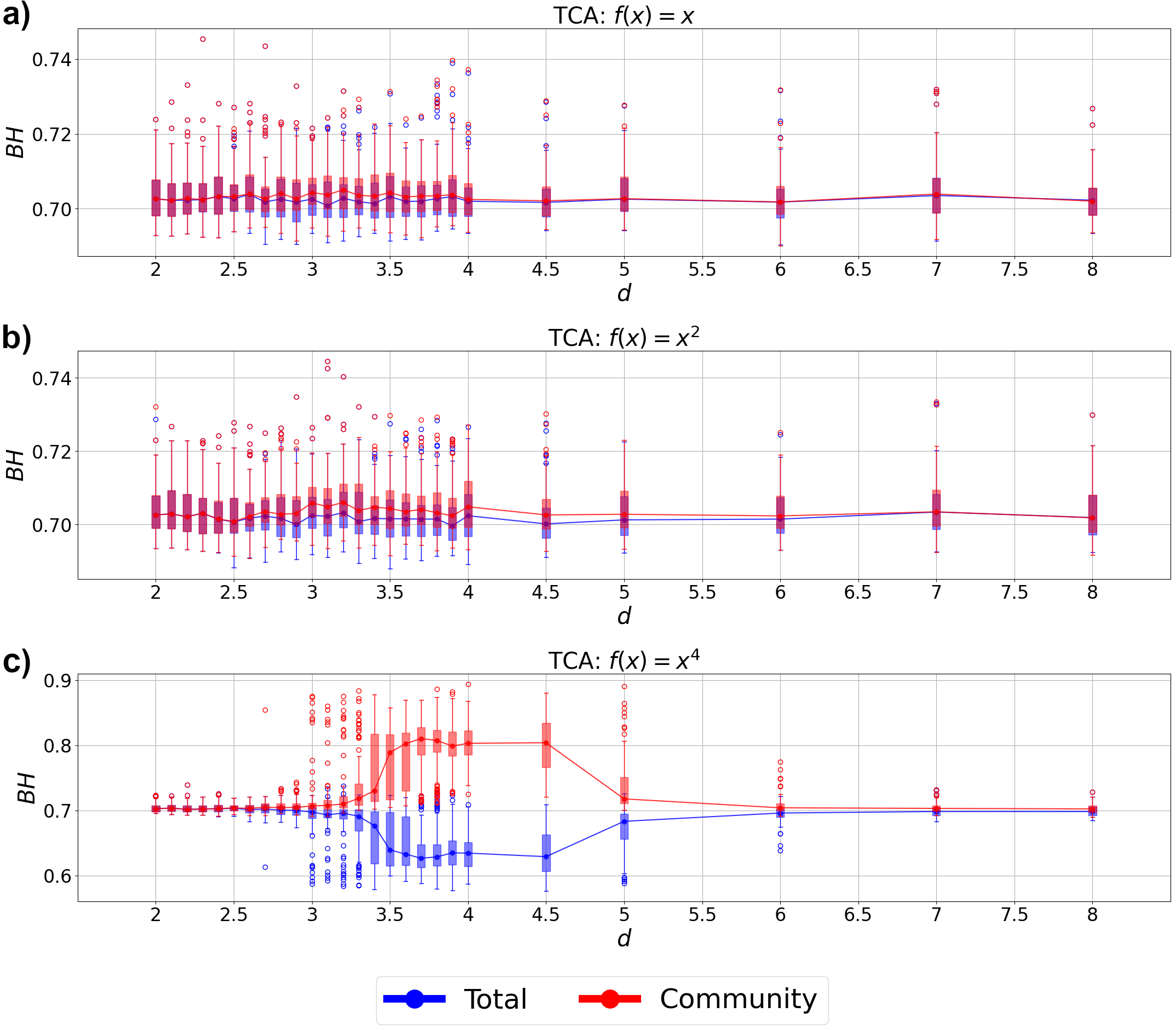}
\caption{\textbf{Belief system homogeneity, $BH$, as function of the dissonance penalty, $d$}. We show the $BH$ calculated based on all the agents in the end state of the simulations in blue, whereas the average $BH$ of communities is shown in red. The symbols represent a standard box plot over 100 runs for each parameter setting according to $d$, and the median values are connected by continuous curves. The applied $TCA$ function is indicated in the panel titles. In panel \textbf{c)}, we can see a separation between the total $BH$ and the community $BH$ values. This coincides with the breakup of the social network in Fig.\ref{fig:combined_community_stats}, when small separated communities begin to form, within which the belief systems of the members become more similar compared to that of agents in other communities.}
\label{fig:combined_belief_homogeneity}
\end{figure}

In Fig.\ref{fig:combined_belief_homogeneity} we show the $BH$ at the end of the simulations among all the agents (blue) together with the average $BH$ obtained for communities (red) as a function of the dissonance penalty $d$ for the different $TCA$ functions. We note that the value of $BH\simeq 0.7$, corresponding to the $BH$ of the entire system at small and large dissonance penalty values (i.e, $d<2.5$ or $6<d$) is slightly higher than the average $BH$ in a null model where the $B_{\alpha,\beta}$ values are independently drawn uniform random variables in the $[0,1]$ interval. According to Fig.\ref{fig:combined_belief_homogeneity}, the $BH$ inside the communities becomes larger compared to the $BH$ in the entire system roughly between $d=3$ and $d=6$, corresponding to the range where the presence of communities in the social network is most significant according to the results described earlier. This effect is most prominent for the simulations using $TCA: f(x)=x^4$ (Fig.\ref{fig:combined_belief_homogeneity}c), where a strong increase in the $BH$ of communities is accompanied by a relevant drop in the $BH$ of the entire system. This shows that the agents in the small communities that appear at this parameter setting develop belief systems that are very similar to the belief system of fellow community members, and in the meantime differ more from the belief system of agents outside the community.    

\section*{Discussion}

In the present paper we have proposed a social network model driven mainly by cognitive dissonance avoidance, corresponding to the the keen effort for maintaining a coherent belief system. In order to study the effects this driving force can have on the structure of the social network, we have used an agent based approach where the agents communicate with each other along the social connections, and are also endowed with realistic human features, most prominently, by a belief system consisting of concepts (beliefs) that are interrelated with one another in an additional internal network. Similar models for the belief systems were already introduced by research communities interested in opinion dynamics, most prominently, but not exclusively\cite{Rodriguez16}, by those aiming to simulate the dynamics of political attitudes\cite{OilSpill, BaldGold14, ConservVsLib, PolBSNWCentral, EvalPolBSNW}. 
In a typical model for the internal network
, the nodes represent "an element of a person's belief system"\cite{EvalPolBSNW}, which are named variously in different models, such as "opinions"\cite{OilSpill}, "concepts"\cite{Rodriguez16} "attitudes"\cite{OilSpill, EvalPolBSNW, PolBSNWCentral}, "beliefs"\cite{ConservVsLib, EvalPolBSNW, Rodriguez16}, or "positions"\cite{PolBSNWCentral, BaldGold14}. 
The links in the network of the belief system correspond to the ties among the concepts, which in the traditional models are usually either positive or negative, depending on their supporting (consistent) or disproving (opposing) nature. 

In contrast with earlier approaches, in the present work we assumed that the links between the attitudes can only be non-negative. 

In parallel, we introduced node weights in the internal network representation of the belief system for representing the attitudes of the agents towards the different concepts, which can be both positive or negative. Accordingly, the association between to concepts can either increase the coherency level of the belief system (providing the pleasant feeling of reassurance when the attitudes towards strongly connected concepts have the same sign), or inversely, it can create cognitive dissonance, which is decreasing the coherence of the belief system. 


We analyzed the behavior of the proposed model through simulations that began with identical agents forming a fully connected social network. As the simulations progressed, both the connections within the social network and the internal belief systems of the agents underwent significant structural changes. The model incorporated two key parameters: the cognitive dissonance penalty, $d$, which represents the strength with which agents strive to avoid cognitive dissonance, and the $TCA$ function, which governs the agents' tendency to initiate communication with the friends of their friends. The simulations were concluded once the social network's structure stabilized. 

According to our results, in this end state of the social network -- depending on the parameter setting -- a wide range of social structures can appear, ranging from consensus through fragmentation to polarization. Consensus, where the social network remains dense, with all the agents forming a single group, is probably the most simple outcome we can expect based on the extremely homogeneous initial state of the simulations. This end state occurs at low values of $d$, where agents can tolerate more the possible cognitive dissonance that may emerge due to the communications between each other. However, at moderate $d$ values and moderately increasing $TCA$ functions (that is $f(x)=x$ or $f(x)=x^2$) the social network becomes polarised, where a small number of large communities form. Members of these communities tend to have more similar belief systems among themselves compared to 'outsiders' both according to the attitudes and the structure of the association weights between the concepts. Last but not least, for a rapidly increasing $TCA$ function of $f(x)=x^4$ we observed fragmentation of the social network in the moderate $d$ regime, where a large number of small communities formed (composed of only a few agents each), that often became isolated from the rest of the network.  These small communities show a larger homogeneity in terms of the belief system of the members compared to the large communities mentioned previously. 

The fact that a drastic change occurred in the structure of the social network roughly around $d\simeq 3$ for all studied $TCA$ functions indicates that the cognitive dissonance avoidance built into our model acts as a very important driving force with a very strong impact on the emerging network structure. Furthermore, from the point of view of the belief systems, our results illustrate that this driving force can also create a wide range of often antagonistic and extreme attitudes even in initially completely homogeneous systems. Naturally, cognitive dissonance avoidance by itself would not induce the above effects for isolated agents. Instead the agents also have to be part of a social network in order to produce these interesting transitions, since it is the communication among the agents that sets any transition process afloat, by changing the strength of associations among the concepts ( -- which is a very basic representation of the mechanism of associative learning). 

Furthermore, here we note that cognitive dissonance can be decreased with various other ways as well (which methods are mastered by humans), for example by focusing on information that reassures the already existing ones while dismissing those that would create "unpleasant" associations -- known as \textit{confirmation bias} --, or by accepting beliefs that "explain away" unpleasant associations (For example, somebody we love did something bad because she/he was forced to). This latter behaviour explains why intelligent people are so prone to accept strongly questionably information so easily, in case it fits well to their belief system. As future directions, the inclusion of such cognitive dissonance decreasing strategies can also be incorporated into the model.

To put our work in a more general context, we refer to the widely shared expectations from just a couple of years ago that various social media platforms would unify public opinion\cite{TwitterIiB}. Currently right the opposite seems to happen, as the prevalence of opinion polarisation in most societies has significantly increased rather than diminished since the emergence of these platforms\cite{USC2024, Vanderbilt2024}. A plausible reason for this is that social media platforms allow their users to connect with others who reinforce their own views, thereby solidifying their echo chambers. Related to that, a straightforward continuation of the present study (planned for future work) is to allow variations among the agents in terms of which are the concepts that have a fixed constant positive or a fixed constant negative attitude in the belief system (representing communities with different value systems). Another continuation can be the study of possible ways by which polarized communities can be depolarized, using the same framework.

\section*{Methods}\label{sec:Methods}






\subsection*{Metrics}

To help monitor the behaviour of our system, we defined several custom metrics. Their detailed explanation can be found below.

\subsubsection*{Extremism}
To measure the extremism, $EM$ of the attitudes of the agents, we can the sum of the absolute attitude values towards each belief, normalized by the belief system size $M$, written as
\begin{equation}
    EM_i = \frac{1}{M}\sum_{\alpha}\left| a^{(i)}_{\alpha}\right|
\label{eq:extremism}
\end{equation}
The higher the extremism, the closer the attitudes are to the two poles of -1 and 1.

\subsubsection*{Attitude homogeneity}
This metric is based on pairwise comparisons of the agents. For every pair, we take the sum of the absolute difference between the non-constant attitudes towards the same concepts. Then to normalize it, we divide by two times the number of non-constant attitudes, $M^*$, so that it has a maximum value of 1 if the two agents have completely opposite opinions, and 0 if they share the same ones. To transform this into a homogeneity metric, we subtract this value from 1, so that 1 means that the two agents have completely homogeneous beliefs. According to that, for a given pair of agents $i$ and $j$ we can write
\begin{equation}
    AH_{ij} = 1 - \frac{1}{2M^*}\sum_{\alpha}\left|a_{\alpha}^{(i)} - a_{\alpha}^{(j)}\right|.
    \label{eq:AH_for_pair}
\end{equation}
For any group of agents, such as a community or the entire social network, the attitude homogeneity, $AH$, is given by the average of $AH_{ij}$ over all the possible agent pairs in the group.  

\subsubsection*{Belief system homogeneity}

To measure the similarity between the structure of the belief systems between the agents, we created a metric we call belief system homogeneity, $BH$, which is similar in nature to the attitude homogeneity. For a pair of agents, here take the sum of the absolute difference between each link weight in the belief system, then  divide by the total number of links, given by $M(M-1)/2$. Following that we subtract the obtained value from 1, written as
\begin{equation}
    BH_{ij} = 1-\frac{2}{M(M-1)}\sum_{\alpha < \beta}\left| B_{\alpha\beta}^{(i)} - B_{\alpha\beta}^{(j)} \right|.
    \label{eq:BH_for_pair}
\end{equation}
This way, a final result of $BH_{ij}=1$ shows identical belief system weights, whereas 0 indicates the opposite. For groups of agents, the belief system homogeneity $BH$ is simply given by the average of the $BH_{ij}$ values over all possible pairs of agents within the group.


\subsubsection*{Communities}
To calculate the number of communities in our social network, we used the Leiden~\cite{Leiden} algorithm optimized for modularity included in the igraph Python package. \cite{igraph} The Leiden algorithm~\cite{Leiden} (widely known as the improved version of the Louvain approach~\cite{Louvain}) works by initially assigning all agents their own communities. The nodes are then moved locally to neighboring communities in a way that maximizes an objective function, in the present case, modularity. Next, these communities are refined, meaning that the nodes inside them are merged into another partition, with the constraint that the new communities must be locally well connected. This refined partition is then aggregated into a new network, on which the same local moving and refinement steps are performed. The process then repeats, starting out from this new aggregated network until no further increase in the objective function is possible.

\subsubsection*{Modularity}
Modularity is a metric used to determine the quality of a given network partition. It's defined as the difference between the fraction of edges found in communities and the expected fraction of edges inside the communities. \cite{Newman-Girvan_modularity} For directed network, this can be written in the following form:
\begin{equation}
    Q = \frac{1}{m}\sum_{ij}\left[ A_{ij} - \gamma\frac{k^{in}_{i}k^{out}_{j}}{m}\right] \delta_{c_{i}, c_{j}}.
    \label{directed_modularity}
\end{equation}
Here, $m$ is the total number of edges in the network, $A_{ij}$ is the adjacency matrix containing the edge weights, $\gamma$ is the resolution parameter, $k^{in}_{i}$ is the sum of incoming edge weights to node $i$, $k^{out}_{j}$ is the sum of outgoing edge weights from node $j$, $\delta$ is the Kronecker symbol, $c_i$ and $c_j$ are the partition of node $i$ and $j$ respectively.

\subsection*{Simulations} \label{simulations_section}
In our simulations, we first created the social network, with the given $N$ and $M$ values, as well as the list of constant positive and negative nodes in the belief system and the temperature constant. Other options include flag whether the social network should initialize one random Agent and create $N$ clones, or create all agents randomly. After this, we repeatedly communicate the agents, stopping the process once the following convergence criteria are met:

\begin{enumerate}
    \item The sum of edge weights in the social network shows no significant changes
    \item The subgroup belief homogeneity shows no significant changes
\end{enumerate}

In each case, we compared the average of these measures in the last 1000 data points, taken at every 100th communication to the previous 1000 data points. For the edge weights, we used constant normalization factors instead of the previous results to calculate the percentage change for faster convergence. To stop the simulation, the convergence criteria had to be fulfilled for $c_{count}$ consecutive checks. The simulation results for 250 agent runs can be found in the Supplementary Information in Figs. S4-S7. These runs provided similar results to those discussed here in the main text. The simulation parameters used during the model were the following:

\begin{table}[ht!]
    \centering
    \begin{tabular}{|c|c|c|c|c|}
    \hline
    Parameter & Values used during simulation \\\hline
    N - social network size & 100 - 250 \\\hline
    M - belief system size & 10 \\\hline
    T - stabilization temperature & 0.01 \\\hline
    b - number of stabilization iterations & 10 \\\hline
    Number of constant attitudes & 1 \\\hline
    \end{tabular}
    \caption{Parameter values used during simulations}
\end{table}

\begin{table}[ht!]
    \centering
    \begin{tabular}{|c|c|c|c|c|c|}
    \hline
    $N$ & $M$ & $TCA$ & $c_{edges}$ & $c_{beliefs}$ & $c_{count}$ \\\hline
    100 & 10 & $f(x) = x$ & $d/5000 < 0.5 \% $ & $d/p < 5 \%$ & 200  \\\hline
    100 & 10 & $f(x) = x^2$ & $d/5000 < 0.25 \%$ & $d/p < 5 \%$ & 200  \\\hline
    100 & 10 & $f(x) = x^4$ & $d/1000 < 0.1 \%$ & $d/p < 5 \% $ & 200   \\\hline
    250 & 10 & $f(x) = x$ & $d/10000 < 0.5 \% $ & $d/p < 5 \% $ & 1000 \\\hline
    250 & 10 & $f(x) = x^2$ & $d/10000 < 0.5 \%$ & $d/p < 5 \% $ & 1000  \\\hline
    250 & 10 & $f(x) = x^4$ & $d/1000 < 0.5 \% $ & $d/p < 5 \%$ & 1000 \\\hline 
    \end{tabular}
    \caption{Convergence parameters used during the simulations. $c_{edges}$ and $c_{beliefs}$ are the convergence criteria for the summed edge weights and community belief homogeneity respectively. $d$ is the absolute difference from the previous 1000 data points and $p$ is the mean of the previous 1000 data points as a normalization factor. $c_{count}$ is the number of consecutive convergence checks required to end the simulation.}
\end{table}




\bibliographystyle{unsrt}
\bibliography{Refek}

\begin{thebibliography}{10}

\bibitem{Laci_revmod}
R.~Albert and A.-L. Barab{\'a}si.
\newblock Statistical mechanics of complex networks.
\newblock {\em Rev. Mod. Phys.}, 74:47--97, 2002.

\bibitem{Dorog_book}
J.~F.~F. Mendes and S.~N. Dorogovtsev.
\newblock {\em Evolution of Networks: From Biological Nets to the Internet and
  WWW}.
\newblock Oxford Univ. Press, Oxford, 2003.

\bibitem{Newman_Barabasi_Watts}
M.~E.~J. Newman, A.-L. Barabási, and D.~J. Watts, editors.
\newblock {\em The Structure and Dynamics of Networks}.
\newblock Princeton University Press, Princeton and Oxford, 2006.

\bibitem{Jari_Holme_Phys_Rep}
Petter Holme and Jari Saramäki.
\newblock Temporal networks.
\newblock {\em Physics Reports}, 519(3):97 -- 125, 2012.
\newblock Temporal Networks.

\bibitem{Vespignani_book}
A.~Barrat, M.~Barthelemy, and A.~Vespignani.
\newblock {\em Dynamical processes on complex networks}.
\newblock Cambridge University Press, Cambridge, 2008.

\bibitem{Connected09}
Nicholas~A. Christakis and James~H. Fowler.
\newblock {\em Connected: The Surprising Power of Our Social Networks and How
  They Shape Our Lives}.
\newblock Little, Brown, 2009.

\bibitem{HomophSoc}
Miller McPherson, Lynn Smith-Lovin, and James~M Cook.
\newblock Birds of a feather: Homophily in social networks.
\newblock {\em Annual Review of Sociology}, 27(1):415--444, 2001.

\bibitem{DeGroot74}
Morris~H. DeGroot.
\newblock Reaching a consensus.
\newblock {\em Journal of the American Statistical Association},
  69(345):118--121, 1974.

\bibitem{Axlrd}
Robert Axelrod.
\newblock {\em The Evolution of Cooperation}.
\newblock Basic Books, 1984.

\bibitem{TCSch78}
Thomas~C. Schelling.
\newblock {\em Micromotives and Macrobehavior}.
\newblock W. W. Norton \& Company; Revised edition, 2006.

\bibitem{SG08}
Serge Galam.
\newblock Sociophysics: A review of galam models.
\newblock {\em International Journal of Modern Physics C}, 19, 2008.

\bibitem{AlapOpDynOf}
Claudio Castellano, Santo Fortunato, and Vittorio Loreto.
\newblock Statistical physics of social dynamics.
\newblock {\em Reviews of Modern Physics}, 81:591--646, 2009.

\bibitem{peralta2022opinion}
Antonio~F. Peralta, János Kertész, and Gerardo Iñiguez.
\newblock Opinion dynamics in social networks: From models to data, 2022.

\bibitem{PerraSciRep}
Nicola Perra and Luis E.~C. Rocha.
\newblock Modelling opinion dynamics in the age of algorithmic personalisation.
\newblock {\em Scientific Reports}, 2019.

\bibitem{Sirbu2017}
Alina S{\^i}rbu, Vittorio Loreto, Vito D.~P. Servedio, and Francesca Tria.
\newblock {\em Opinion Dynamics: Models, Extensions and External Effects},
  pages 363--401.
\newblock Springer International Publishing, Cham, 2017.

\bibitem{ContOpDynSurvey}
Jan Lorenz.
\newblock Continuous opinion dynamics under bounded confidence: A survey.
\newblock {\em International Journal of Modern Physics C}, 18(12):1819--1838,
  2007.

\bibitem{SGRev}
Serge Galam.
\newblock {Sociophysics: A Review Of Galam Models}.
\newblock {\em International Journal of Modern Physics C (IJMPC)},
  19(03):409--440, 2008.

\bibitem{CsanyiHied}
Bal\'{a}zs T\'{o}th and Vilmos Cs\'{a}nyi.
\newblock {\em Our Beliefs -- The building blocks of human thoughts (in
  Hungarian: Hiedelmeink -- Az emberi gondolatok \'{e}p\'{i}t\H{o}k\"{o}vei)}.
\newblock Libri, Budapest, 2017.

\bibitem{Scout}
Julia Galef.
\newblock {\em The Scout Mindset: Why Some People See Things Clearly and Others
  Don't}.
\newblock Portfolio, 2021.

\bibitem{KnowldgIllsn}
Steven Sloman and Philip Fernbach.
\newblock {\em The Knowledge Illusion: Why We Never Think Alone}.
\newblock Riverhead books, New York, NY, 2017.

\bibitem{BelBrain}
Michael Shermer.
\newblock {\em The Believing Brain: From Ghosts and Gods to Politics and
  Conspiracies - How We Construct Beliefs and Reinforce Them as Truths}.
\newblock St. Martin's Griffin; Illustrated edition, 2012.

\bibitem{CogPsyTK}
Kathleen~M. Galotti.
\newblock {\em Cognitive Psychology In and Out of the Laboratory}.
\newblock SAGE Publications, Fifth edition, 2013.

\bibitem{SelfishGene}
Richard Dawkins.
\newblock {\em The Selfish Gene}.
\newblock Oxford University Press, 1976.

\bibitem{TC1998}
Aaron Lynch.
\newblock {\em Thought Contagion}.
\newblock Basic Books, 1998.

\bibitem{Converse1964}
Philip~E. Converse.
\newblock The nature of belief systems in mass publics.
\newblock {\em Critical Review}, 18, 1964.

\bibitem{Rokeach63}
Milton Rokeach.
\newblock The organization and modification of beliefs.
\newblock {\em The Centennial Review}, 7:375--395, 1963.

\bibitem{g11040065}
Michel Grabisch and Agnieszka Rusinowska.
\newblock A survey on nonstrategic models of opinion dynamics.
\newblock {\em Games}, 11(4), 2020.

\bibitem{galam1982sociophysics}
Serge Galam, Yuval Gefen, and Yonathan Shapir.
\newblock Sociophysics: A new approach of sociological collective behaviour. i.
  mean-behaviour description of a strike.
\newblock {\em Journal of Mathematical Sociology}, 9(1):1--13, 1982.

\bibitem{stone1961opinion}
Mervyn Stone.
\newblock The opinion pool.
\newblock {\em The Annals of Mathematical Statistics}, pages 1339--1342, 1961.

\bibitem{chatterjee1977towards}
Samprit Chatterjee and Eugene Seneta.
\newblock Towards consensus: Some convergence theorems on repeated averaging.
\newblock {\em Journal of Applied Probability}, 14(1):89--97, 1977.

\bibitem{8ae4e1e508ad490382855c2970e950b7}
Approaching consensus can be delicate when positions harden.
\newblock {\em Stochastic Processes and their Applications}, 22(2):315--322,
  July 1986.

\bibitem{Olle14}
Olle H\"{a}ggstr\"{o}m and Timo Hirscher.
\newblock Further results on consensus formation in the deffuant model.
\newblock {\em Electronic Journal of Probability}, 19:26, 2014.

\bibitem{OrigDeffuant}
Guillaume Deffuant, David Neau, Frederic Amblard, and Gérard Weisbuch.
\newblock Mixing beliefs among interacting agents.
\newblock {\em Advances in Complex Systems}, 03(01n04):87--98, 2000.

\bibitem{AxelrodUjPolar}
Robert Axelrod, Joshua~J. Daymude, and Stephanie Forrest.
\newblock Preventing extreme polarization of political attitudes.
\newblock {\em Proceedings of the National Academy of Sciences},
  118(50):e2102139118, 2021.

\bibitem{Latan1981ThePO}
Bibb Latan{\'e}.
\newblock The psychology of social impact.
\newblock {\em American Psychologist}, 36:343--356, 1981.

\bibitem{AxelrodCikk}
Robert Axelrod.
\newblock The dissemination of culture: A model with local convergence and
  global polarization.
\newblock {\em The Journal of Conflict Resolution}, 41(2):203--226, 1997.

\bibitem{RADUCHA2017427}
Tomasz Raducha and Tomasz Gubiec.
\newblock Coevolving complex networks in the model of social interactions.
\newblock {\em Physica A: Statistical Mechanics and its Applications},
  471:427--435, 2017.

\bibitem{AssocLBk}
Stephen~B. Klein.
\newblock {\em Learning: Principles and Applications}.
\newblock SAGE Publications, Inc; 8th edition, 2018.

\bibitem{AssocL19}
Simon De~Deyne, Danielle~J. Navarro, Amy Perfors, Marc Brysbaert, and Gert
  Storms.
\newblock The “small world of words” english word association norms for
  over 12,000 cue words.
\newblock {\em Behavior Research Methods}, 2019.

\bibitem{8945152}
Zheng Hu, Jiao Luo, Chunhong Zhang, and Wei Li.
\newblock A natural language process-based framework for automatic association
  word extraction.
\newblock {\em IEEE Access}, 8:1986--1997, 2020.

\bibitem{CDL:Cooper07}
Joel Cooper.
\newblock {\em Cognitive Dissonance: 50 Years of a Classic Theory}.
\newblock SAGE Publications Ltd, 2007.

\bibitem{CDL:MedicalNewsT}
Jayne Leonard.
\newblock Cognitive dissonance: What to know.
\newblock \url{https://www.medicalnewstoday.com/articles/326738}, 2019.

\bibitem{RepulsiveEffect}
Christopher~A. Bail, Lisa~P. Argyle, Taylor~W. Brown, John~P. Bumpus, Haohan
  Chen, M.~B.~Fallin Hunzaker, Jaemin Lee, Marcus Mann, Friedolin Merhout, and
  Alexander Volfovsky.
\newblock Exposure to opposing views on social media can increase political
  polarization.
\newblock {\em Proceedings of the National Academy of Sciences},
  115(37):9216--9221, 2018.

\bibitem{NoDistancing}
Michael M\"as and Andreas Flache.
\newblock Differentiation without distancing. explaining bi-polarization of
  opinions without negative influence.
\newblock {\em PLoS ONE}, 8(11):e74516, 2013.

\bibitem{EBGoldsteinCognPsy}
E.~Bruce Goldstein.
\newblock {\em Cognitive Psychology: Connecting Mind, Research and Everyday
  Experience}.
\newblock Cengage Learning; 5th edition, 2018.

\bibitem{HeiderBT}
Fritz Heider.
\newblock Attitudes and cognitive organization.
\newblock {\em The Journal of Psychology}, 21:107--112, 1946.

\bibitem{ESTRADA201970}
Ernesto Estrada.
\newblock Rethinking structural balance in signed social networks.
\newblock {\em Discrete Applied Mathematics}, 268:70--90, 2019.

\bibitem{DoB}
Talaga Szymon, Massimo Stella, Trevor~James Swanson, and Andreia~Sofia
  Teixeira.
\newblock Polarization and multiscale structural balance in signed networks.
\newblock {\em Communications Physics}, 6, 2023.

\bibitem{Rodriguez16}
Nathaniel Rodriguez, Johan Bollen, and Yong-Yeol Ahn.
\newblock Collective dynamics of belief evolution under cognitive coherence and
  social conformity.
\newblock {\em PlosOne}, 11(11), 2016.

\bibitem{PhysRevE.90.042802}
Ernesto Estrada and Michele Benzi.
\newblock Walk-based measure of balance in signed networks: Detecting lack of
  balance in social networks.
\newblock {\em Phys. Rev. E}, 90:042802, Oct 2014.

\bibitem{moth1}
John Bowlby.
\newblock {\em Attachment and Loss, Vol. 1: Attachment}.
\newblock Basic Books, 1969.

\bibitem{moth2}
Robert~A. LeVine, Patrice~M. Miller, and Mary~Maxwell West, editors.
\newblock {\em Parental Behavior in Diverse Societies}.
\newblock Jossey-Bass, 1988.

\bibitem{moth3}
Nick~J. Royle, Per~T. Smiseth, and Mathias Kölliker.
\newblock {\em {The Evolution of Parental Care}}.
\newblock Oxford University Press, 08 2012.

\bibitem{fear}
Joanna Bourke.
\newblock {\em Fear: A Cultural History}.
\newblock Virago, 2006.

\bibitem{death}
Ernest Becker.
\newblock {\em The Denial of Death}.
\newblock Free Press, 1997.

\bibitem{CultDiff}
Geert Hofstede.
\newblock {\em Culture's Consequences}.
\newblock Sage Publications, 1984.

\bibitem{CognDissA}
Eddie Harmon-Jones, editor.
\newblock {\em Cognitive Dissonance: Reexamining a Pivotal Theory in
  Psychology}.
\newblock American Psychological Association; 2nd edition, 2019.

\bibitem{Watts1998}
Duncan~J. Watts and Steven~H. Strogatz.
\newblock Collective dynamics of `small-world' networks.
\newblock {\em Nature}, 393(6684):440--442, Jun 1998.

\bibitem{triadic_closure}
David Easley and Jon Kleinberg.
\newblock {\em Networks, Crowds, and Markets: Reasoning About A Highly
  Connected World}.
\newblock Cambridge University Press, 07 2010.

\bibitem{Leiden}
V.~Traag, L.~Waltman, and Nees~Jan van Eck.
\newblock From louvain to leiden: guaranteeing well-connected communities.
\newblock {\em Scientific Reports}, 9:5233, 03 2019.

\bibitem{Fortunato_coms}
Santo Fortunato.
\newblock Community detection in graphs.
\newblock {\em Physics Reports}, 486(3):75 -- 174, 2010.

\bibitem{Fortunato_Hric_coms}
Santo Fortunato and Darko Hric.
\newblock Community detection in networks: A user guide.
\newblock {\em Physics Reports}, 659:1 -- 44, 2016.
\newblock Community detection in networks: A user guide.

\bibitem{Cherifi_coms}
H.~Cherifi, G.~Palla, B.K. Szymanski, and X.~Lu.
\newblock On community structure in complex networks: challenges and
  opportunities.
\newblock {\em Appl. Netw. Sci.}, 4:117, 2019.

\bibitem{Newman-Girvan_modularity}
M.~E.~J. Newman and M.~Girvan.
\newblock Finding and evaluating community structure in networks.
\newblock {\em Phys. Rev. E}, 69:026113, Feb 2004.

\bibitem{tsne}
Laurens van~der Maaten and Geoffrey Hinton.
\newblock Visualizing data using t-sne.
\newblock {\em Journal of Machine Learning Research}, 9(86):2579--2605, 2008.

\bibitem{umap}
Leland McInnes, John Healy, Nathaniel Saul, and Lukas Großberger.
\newblock Umap: Uniform manifold approximation and projection.
\newblock {\em Journal of Open Source Software}, 3(29):861, 2018.

\bibitem{OilSpill}
Daniel DellaPosta.
\newblock Pluralistic collapse: The “oil spill” model of mass opinion
  polarization.
\newblock {\em American Sociological Review}, 85(3):507--536, 2020.

\bibitem{BaldGold14}
Delia Baldassarri and Amir Goldberg.
\newblock Neither ideologues nor agnostics: Alternative voters' belief system
  in an age of partisan politics.
\newblock {\em American Journal of Sociology}, 120(1):45--95, 2014.

\bibitem{ConservVsLib}
Felicity~M. Turner-Zwinkels, Branden~B. Johnson, Chris~G. Sibley, and Mark~J.
  Brandt.
\newblock Conservatives' moral foundations are more densely connected than
  liberals' moral foundations.
\newblock {\em Personality and Social Psychology Bulletin}, 47(2):167--184,
  2020.

\bibitem{PolBSNWCentral}
Mark~J. Brandt, Chris~G. Sibley, and Danny Osborne.
\newblock What is central to political belief system networks?
\newblock {\em Personality and Social Psychology Bulletin}, 45(9):1352--1364,
  2019.

\bibitem{EvalPolBSNW}
Mark~J. Brandt and Willem W.~A. Sleegers.
\newblock Evaluating belief system networks as a theory of political belief
  system dynamics.
\newblock {\em Personality and Social Psychology Review}, 25(2):159--185, 2021.

\bibitem{TwitterIiB}
David Streitfeld.
\newblock The internet is broken.
\newblock {\em The New York Times}, 2017.

\bibitem{USC2024}
{USC Annenberg School for Communication and Journalism}.
\newblock Usc polarization index reveals america’s political divide remains
  high.
\newblock
  \url{https://annenberg.usc.edu/news/usc-polarization-index-reveals-americas-political-divide-remains-high},
  2024.
\newblock Accessed: 2024-08-19.

\bibitem{Vanderbilt2024}
{Vanderbilt University}.
\newblock Latest vanderbilt unity index shows the u.s. continuing its trend
  toward increased political polarization.
\newblock
  \url{https://news.vanderbilt.edu/2024/02/14/latest-vanderbilt-unity-index-shows-the-u-s-continuing-its-trend-toward-increased-political-polarization/},
  2024.
\newblock Accessed: 2024-08-19.

\bibitem{igraph}
Gabor Csardi and Tamas Nepusz.
\newblock The igraph software package for complex network research.
\newblock {\em InterJournal}, Complex Systems:1695, 11 2005.

\bibitem{Louvain}
Vincent~D Blondel, Jean-Loup Guillaume, Renaud Lambiotte, and Etienne Lefebvre.
\newblock Fast unfolding of communities in large networks.
\newblock {\em Journal of Statistical Mechanics: Theory and Experiment},
  2008(10):P10008, oct 2008.

\end{thebibliography}

\section*{Acknowledgements}

This research was partially supported by the National Research, Development and Innovation Office – NKFIH, grant No. K128780.


\section*{Author contributions}

A. Z. and G. P. developed the concept of the study, Z. K. implemented the model, carried out the simulations and performed the analyses. Z. K. prepared the figures, G. P. and A. Z. wrote the paper. All authors reviewed the manuscript. 


\section*{Data availability}

All data generated during the current study are available from the corresponding author upon request.

\section*{Code availability}
The code used for the simulations is available at \href{https://github.com/ztoli17/Assoc-based-bsdyn}{https://github.com/ztoli17/Assoc-based-bsdyn}

\newpage

\begin{center}
\LARGE{\bf SUPPLEMENTARY INFORMATION}
\end{center}

\renewcommand{\thefigure}{S\arabic{figure}}
\renewcommand{\thetable}{S\arabic{table}}
\renewcommand{\theequation}{S\arabic{equation}}
\renewcommand{\thesection}{S\arabic{section}}

\setcounter{section}{0}
\setcounter{figure}{0}
\setcounter{equation}{0}
\setcounter{table}{0}

\section{Flowcharts of  simulation process}
\begin{figure}[ht!]
    \centering
    \includegraphics[width=0.65\linewidth]{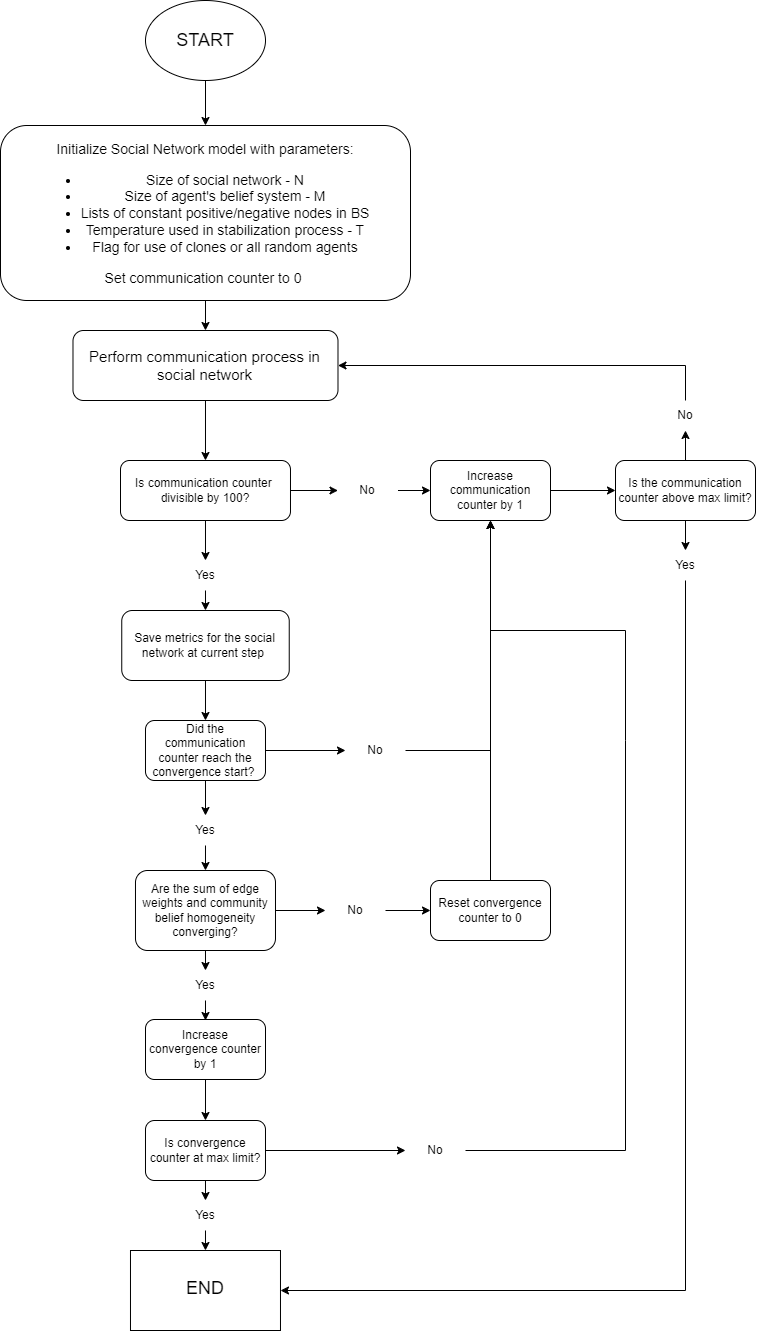}
    \caption{Flowchart detailing the simulation process in the social network}
    \label{simulation_flowchart}
\end{figure}

\begin{figure}[ht!]
    \centering
    \includegraphics[width=0.65\linewidth]{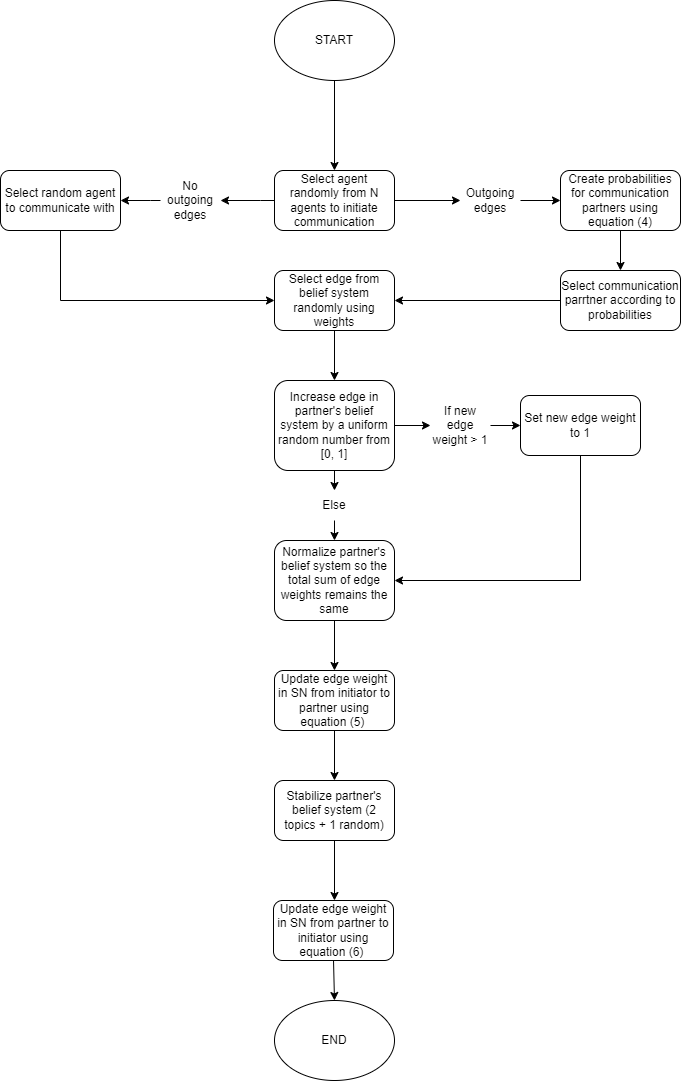}
    \caption{Flowchart detailing the communication process in the social network}
    \label{comm_flowchart}
\end{figure}

\begin{figure}[ht!]
    \centering
    \includegraphics[width=0.65\linewidth]{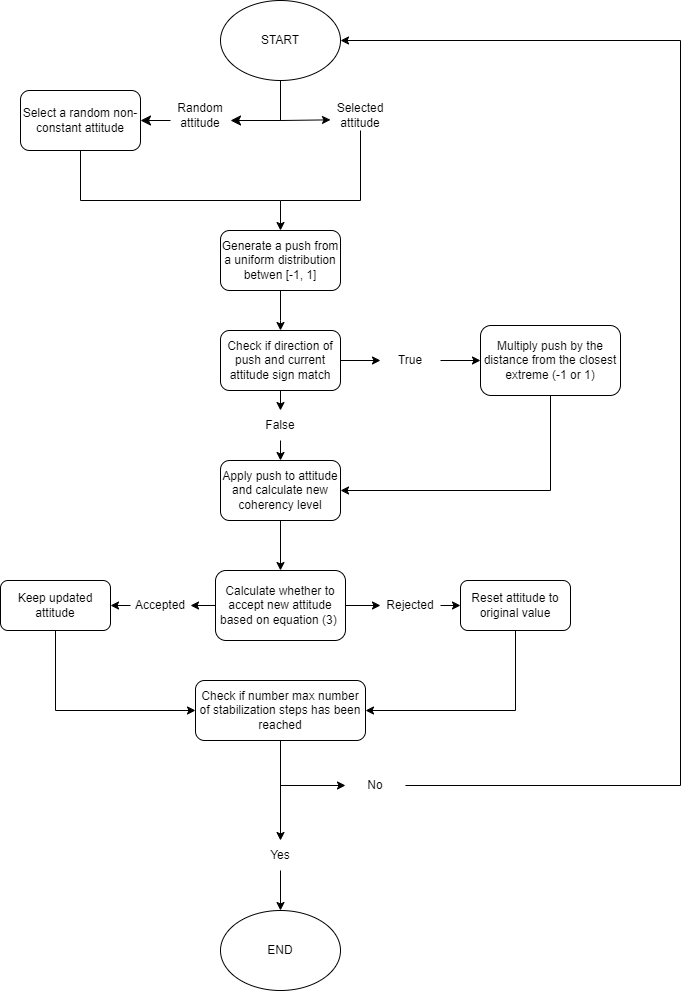}
    \caption{Flowchart detailing the stabilization process of agents}
    \label{stabil_flowchart}
\end{figure}

To supplement the model description, we present also flowcharts depicting the logic of the simulations used in the main paper. Fig.\ref{simulation_flowchart} shows the flowchart detailing the simulation process in the social network, Fig.\ref{comm_flowchart} shows the flowchart detailing the communication process in the social network and Fig. \ref{stabil_flowchart} shows the flowchart detailing the stabilisation process of agents.

\clearpage

\section{250 agent simulation results}

The figures below show the results of simulations with 250 agents. Due to the larger size of the social network, the run-time of the simulations was longer, so we used fewer dissonance penalty values. The results show similar behaviour to those obtained on 100 agent simulations. 

\begin{figure}[h!]
\centering
\includegraphics[width=\linewidth]{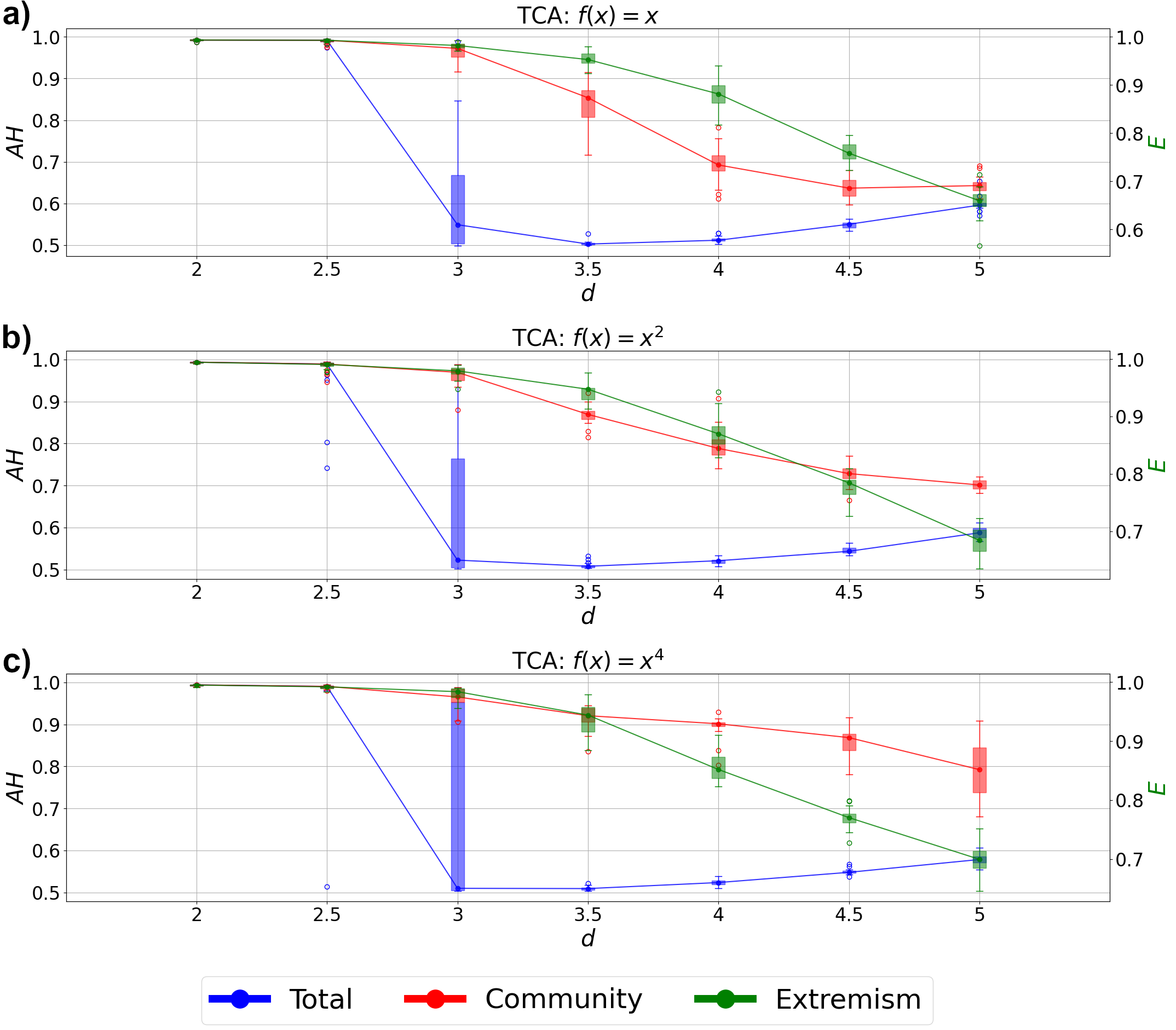}
\caption{The total and intra-community homogeneity of the agent's attitudes. While the total homogeneity drops sharply around a dissonance penalty of 3, the communities stay close in attitudes. This shows that in our model, the agents try to form social groups with others who have similar attitudes towards most topics. This figure is analogous to Fig.4 in the main text.}
\label{combined_attitude_homogeneity}
\end{figure}

\begin{figure}[h!]
\centering
\includegraphics[width=\linewidth]{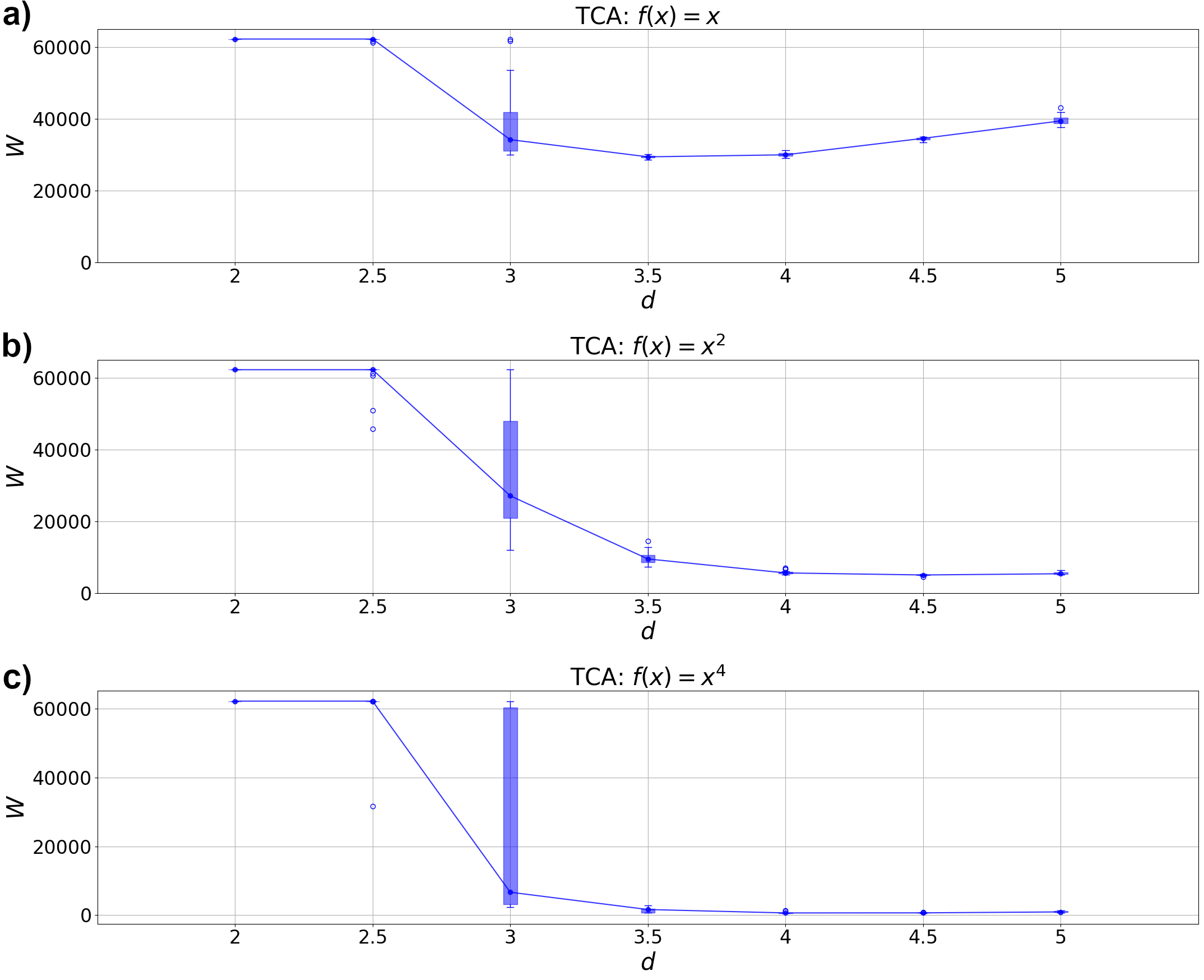}
\caption{Sum of edge weights after convergence for different dissonance penalties. We can see that for the $f(x) = x$ case around half of the possible edges still remain, but for the two stricter $TCA$-s, a vast majority of the edges disappear, resulting in a sparse social network. This figure is analogous to Fig.5 in the main text.}
\label{combined_edge_weights}
\end{figure}

\begin{figure}[h!]
\centering
\includegraphics[width=\linewidth]{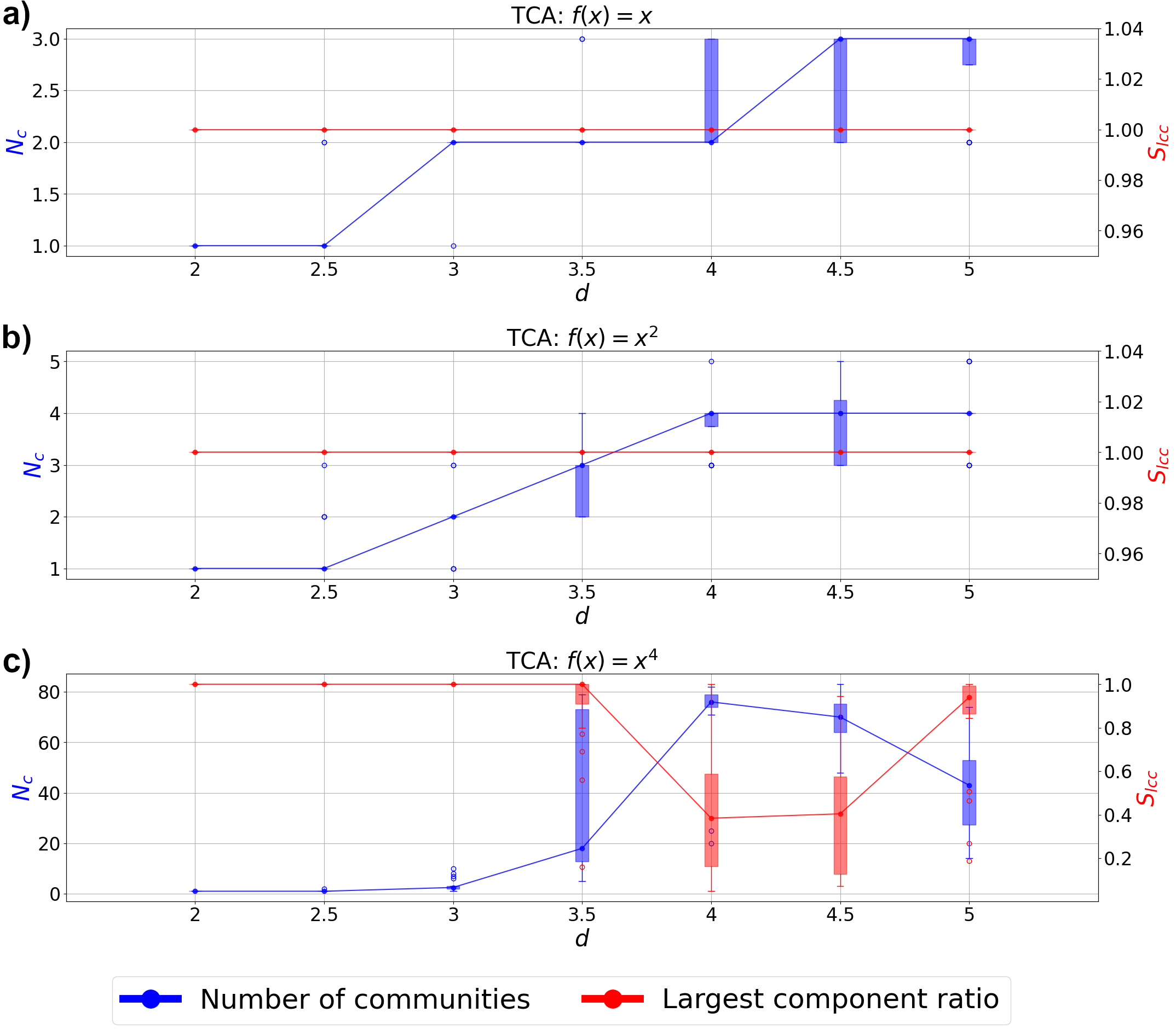}
\caption{The number of communities and the ratio of the largest component to the total nodes in the social network. The fully connected social network starts to break down at around a dissonance penalty of 3, which results in on average 2 groups for $f(x) = x$ $TCA$, and 20 to 30 for the stricter $TCA$-s. This simultaneously means that the giant component slowly disappears, and the communities become isolated from each other. This figure is analogous to Fig.6 in the main text.}
\label{combined_community_stats}
\end{figure}

\begin{figure}[h!]
\centering
\includegraphics[width=\linewidth]{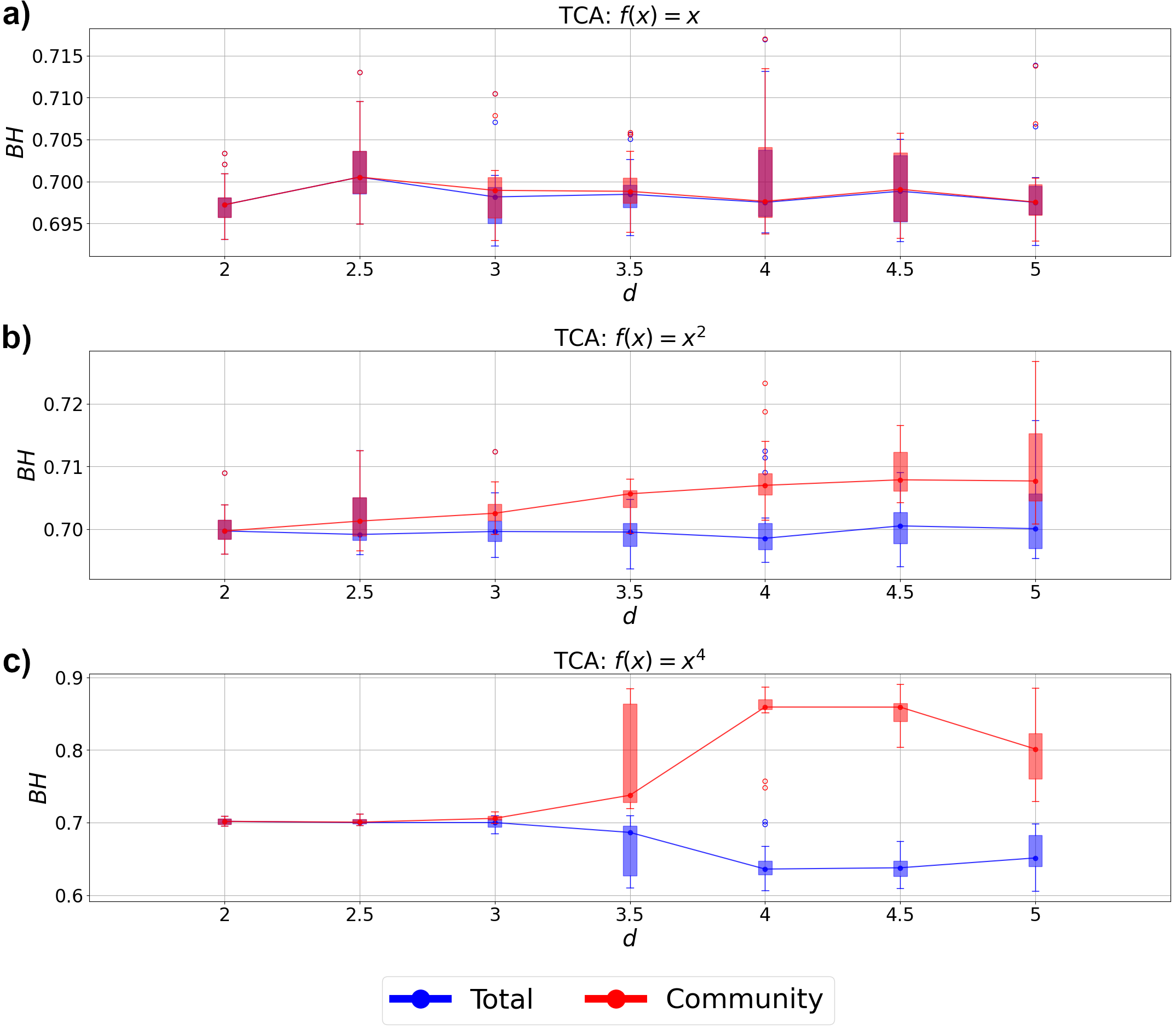}
\caption{The total and intra-community homogeneity of the belief systems. We can also see the difference between the $TCA$: $f(x) = x$ case, and the stricter $TCA$-s. When the social network becomes sparse, the belief systems inside the communities also become more similar. This figure is analogous to Fig.8 in the main text.}
\label{combined_belief_homogeneity}
\end{figure}

\end{document}